\documentclass[a4paper,fleqn]{mnras}

\usepackage[T1]{fontenc}
\usepackage{ae,aecompl}

\usepackage{graphicx}	
\usepackage{amsmath}	
\usepackage{amssymb}	
\usepackage{longtable}
\usepackage{multicol}

\title[A pulsar-runaway-pair that ejected $^{60}$Fe]
{A nearby recent supernova that ejected the runaway star $\zeta$~Oph,
the pulsar PSR\,B1706-16, and $^{60}$Fe found on Earth}

\author[R. Neuh\"auser et al.]{
R. Neuh\"auser,$^{1}$\thanks{E-mail: rne@astro.uni-jena.de},
F. Gie\ss{}ler$^{1}$,
and V.V. Hambaryan$^{1,2}$
\\
$^{1}$Astrophysikalisches Institut und Universit\"ats-Sternwarte Jena, Schillerg\"a\ss{}chen 2-3, 07745 Jena, Germany \\
$^{2}$Byurakan Astrophysical Observatory, Byurakan 0213, Aragatzotn, Armenia
}

\date{Accepted XXX. Received XXX; in original form July 2019}

\pubyear{2019}

\begin{document}
\label{firstpage}
\pagerange{\pageref{firstpage}--\pageref{lastpage}}
\maketitle

\begin{abstract}
The detection of $\sim 1.5-3.2$ Myr old $^{60}$Fe on Earth indicates 
recent nearby core-collapse supernovae. For supernovae in multiple stars, the 
primary stars may become neutron stars, while former companions may become
unbound and become runaway stars. 
We wrote software for tracing back the space motion of runaway and neutron stars 
to young associations of massive stars.
We apply it here to the nearby young Scorpius-Centaurus-Lupus groups,
all known runaway stars possibly coming from there, 
and all 400 neutron stars with known transverse velocity.
We find kinematic evidence that the runaway 
$\zeta$ Oph and the radio pulsar PSR\,B1706-16 were released by a 
supernova in a binary $1.78 \pm 0.21$ Myr ago at $107 \pm 4$ pc 
distance (for pulsar radial velocity $260 \pm 43$ km/s); association age and flight 
time determine the progenitor mass (16-18 M$_{\odot}$), which can constrain supernova 
nucleosynthesis yields and $^{60}$Fe uptake on Earth. In addition, we 
notice that the only high-mass X-ray binary in Scorpius-Centaurus-Lupus (1H11255-567 
with $\mu^{1}$ and $\mu^{2}$ Cru) may include a neutron star formed 
in another SN, up to $\sim 1.8$ Myr ago at $89-112$ pc, i.e. also yielding 
$^{60}$Fe detectable on Earth. Our scenario links $^{60}$Fe found on 
Earth to one or two individual supernovae in multiple stars.
\end{abstract}




\section{Introduction: $^{60}$Fe on Earth}

There is ample experimental evidence for $\sim 1.5-3.2$ Myr old live $^{60}$Fe on Earth, 
produced by recent nearby core-collapse supernovae (SNe) of massive stars 
(Knie et al. 1999, Fry et al. 2015, Wallner et al. 2016). 
Given the time distribution of the detected $^{60}$Fe, it is very likely that
several SNe were responsible (Wallner et al. 2016, their figure S6).
If this $^{60}$Fe can be linked to individual SNe with known distances, ages, and progenitor masses, 
SN nucleosynthesis models and $^{60}$Fe uptake on Earth can be constrained 
(Fry et al. 2015; Breitschwerdt et al. 2016, henceforth B16). 
While theoretical expectations exist for distances and times of several SNe
that would have ejected the $^{60}$Fe detected on Earth, also the masses of their progenitor stars, 
and $^{60}$Fe yields (Fry et al. 2015, 2016, B16, Hyde \& Pecaut 2018), we approach this problem empirically 
by linking such SNe to their other products, namely tracing back neutron stars and runaway stars: 
since most massive stars are in multiple systems, SNe do not only form neutron stars 
(and $^{60}$Fe), but often also eject former companions, which run away with their 
orbital velocity (Blauuw 1961, Poveda et al. 1967, Renzo et al. 2019): 
the smaller the former separation and the higher the masses, the faster. 
 
For calculating backward the three-dimensional (3D) flight path, celestial coordinates, 
distance, and 2D proper motion of runaways and pulsar are known, 
also radial velocities (RV) of runaways, but RVs of pulsars are often unknown. 
Hence, we draw the RV from their most likely Maxwellian distribution of
the space velocities in a Monte Carlo simulation 
(Hoogerwerf et al. 2000, 2001, henceforth Ho00 \& Ho01, respectively; 
Hobbs et al. 2005, Tetzlaff et al. 2010, Verbunt et al. 2017).

A first attempt to trace back runaway stars and neutron stars was performed by Ho00 \& Ho01
and then applied to Hipparcos data: 
they claimed that the runaway star $\zeta$ Oph and the radio pulsar PSR\,B1929+10
were ejected by a SN in a binary $\sim 1$ Myr ago, from the Upper-Scorpius (UpSco)
subgroup of the Scorpius-Centaurus-Lupus association (Ho00, Ho01).
This solution was doubted when better astrometric data became available 
(Chatterjee et al. 2004, Tetzlaff et al. 2010, Kirsten et al. 2015).
Other cases, where a neutron star was traced back to an OB association,
with or without a runaway star, were presented by Tetzlaff et al. 
(2009, 2010, 2011b, 2012, 2013, 2014a, 2014b).

That the binary SN ejection mechanism for producing runaway stars works, 
can also be shown inside SN remnants:
E.g., $\sim 30$ kyr ago, both a pulsar and a runaway star were 
near the center of the SN remnant S147 (Din{\c c}el et al. 2015);
the distance of SN and SNR was estimated from the
spectrophotometric distance towards the runaway star HD 37424 ($1.3 \pm 0.1$ kpc),
confirmed by Gaia DR2 ($1.47 \pm 0.15$ kpc).
Most recently numerical simulations of the evolution of binary stars with one supernova progenitor
were performed by Zapartas et al. (2017b) and Renzo et al. (2019), 
who could show that about one third of original binaries
should produce at least one ejected runaway star.

We developed software to trace back the motion of stars through the Galactic potential (Sect. 2) 
and tested it successfully with the original input data of the first runaway-pulsar-pair,
$\zeta$ Oph and PSR\,B1929+10 (Ho00, Ho01), see our Sect. 3.
We then applied our software to their most recent data showing that $\zeta$ Oph and PSR\,B1929+10 
did not have a common origin, so that other solutions have to be sought. 
Since the Gaia satellite recently determined the astrometric properties (distance, proper motion, RV) 
of many stars and associations (Gaia 2018), we searched more generally for runaway-pulsar-systems that 
were ejected from SNe in Scorpius-Centaurus-Lupus, the most nearby young association of massive stars
(Sect. 4).
While Tucana-Horologium (Tuc-Hor) has also been suggested as possible site of a nearby recent SN 
(Mamajek 2016,
Hyde \& Pecaut 2018), 
Scorpius-Centaurus-Lupus is a priori more likely, because it contains many more massive stars.

Several SNe from Scorpius-Centaurus-Lupus should have produced runaways, neutron stars and $^{60}$Fe 
in the last few Myr (Maiz-Apellaniz 2001): by comparing the current mass functions of the nearby ($\sim 115-136$ pc) 
Upper-Centaurus-Lupus (UCL) and Lower-Centaurus-Crux (LCC) sub-groups of the Scorpius-Centaurus-Lupus 
association (Wright \& Mamajek 2018) with the initial mass function (IMF), $\sim 10$ massive stars are considered 
missing from UCL, $\sim 6$ from LCC (Fuchs et al. 2006, B16). 
The IMF predicts the most likely main-sequence masses of the $\sim 16$ missing 
stars: $\sim 8.8-19.9$ M$_{\odot}$ (B16). 
We discuss the B16 model in more detail in Sect. 5.

If the pulsar beaming fraction is $\sim 1/8$ (Kolonk et al. 2004) and if $\sim 1/3$ of massive stars release runaway stars 
in core-collapse SNe (Renzo et al. 2019, Zapartas et al. 2017b), 
then, among $\sim 16$ expected SNe (B16),
we would expect about one detectable pulsar-runaway-pair --
and we did find one case (Sect. 4). 
In another third, the neutron star remains bound after the SN, sometimes detectable 
as low- or high-mass X-ray binary: we may have found one such case in LCC, 
where a SN would have happened within the last $\sim 1.8$ Myr at $89-112$ pc (Sect. 6).
We finish with a discussion in Sect. 7 and a summary of results in Sect. 8.

Live $^{60}$Fe ($\sim 2$ Myr old, half-life 2.6 Myr) was found not only on Earth 
in deep ocean ferromanganese crusts, nodules, and sediments
(Knie et al. 1999, Fry et al. 2015, Wallner et al. 2016),
and with much less time resolution also in Antarctic ice (Koll et al. 2019),
and in lunar regolith (Firmani et al. 2016),
but also in Galactic cosmic rays with the Advanced Composition Explorer
indicating at least two SNe within $\sim 620$ pc in the last several Myr (Binns et al. 2016).
Further evidence for a local SN $\sim 2-3$ Myr ago was found in otherwise
unexplained features of cosmic ray spectra of nuclei (Savchenko et al. 2015, Kachelrie\ss~et al. 2018).
Lopez-Coto et al. (2018) then presented evidence for a young pulsar ($\sim 10^{5-6}$ yr)
within $\sim 30-80$ pc of Earth
(otherwise unknown, not beaming towards us), which could be the source
for the local high-energy cosmic ray all-electron spectrum, in particular for the positron excess,
because of the low diffusion coefficient (Lopez-Coto et al. 2018).
A possible neutron star within 100 pc (not beaming towards us) could be the companion candidate near Fomalhaut,
which was shown to possibly be a young (few Myr) neutron star ($\sim 10^{5}$ K) in the background of Fomalhaut (i.e. $\ge 11$ pc),
which would have been born within $\sim 72$ pc of Earth, if $\sim 2$ Myr old and if now at $\sim 11$ pc
(Neuh\"auser et al. 2015).
Deep X-ray observations with Chandra of the faint object near Fomalhaut failed to detect it, 
but placed upper limits such that it can still be a neutron star that is cooler than
91000 K and closer than 13.5 pc (Poppenhaeger et al. 2017).

\section{Software to trace back stars through the Galactic potential}

We developed software to trace back the motion of stars through the Galactic potential.
To trace back a star's orbit (position vector $\vec{r}$), 
the well-known equation of motion
\begin{equation}
\ddot{\vec{r}} = -\vec{\nabla} \Phi
\end{equation}
is numerically integrated backwards in time for the Galactic potential $\Phi$. 
The number of simulated orbits per star (here: 3 million) 
as well as the number of steps and step width (here: 1000 years) for the integration are configurable. 
After each step the mutual separations between the two objects,
e.g. runaway star and pulsar, are calculated and the minimum separations are determined.

The Galactic potential $\Phi$ used is a three-component model
\begin{equation}
\Phi = \Phi _{disk} + \Phi _{bulge} + \Phi _{halo}  
\end{equation} 
consisting of potentials for disk (d), spheroid bulge (b), and halo (h) (Johnston et al. 1996),
quite similar -- if not the same -- as used by Ho00 and Ho01: 
\begin{displaymath}
\Phi _{1} = - \frac{G \cdot M_{d}}{\sqrt{x^2 + y^2 + \left(a_d + \sqrt{z^2 + b_d^2}\right)^2}} 
\end{displaymath}
\begin{equation}
- \frac{G \cdot M_{b}}{c + \sqrt{x^2+y^2+z^2}} + v_{h}^2 \cdot \ln\left(x^2+y^2+z^2+d^2\right)
\end{equation}
with an axisymmetric disk as first term (Miyamoto \& Nagai 1975),
a spherically symmetric bulge as second term (Hernquist 1990), 
and a massive, truncated spherical Galactic halo as third term (Johnston et al. 1996); 
we used masses M$_{d}=1.0 \cdot 10^{11}$\,M$_\odot$ and M$_{b}=3.4 \cdot 10^{10}$\,M$_\odot$, 
velocity $v_{h}=128$\,km\,s$^{-1}$, 
and scale parameters $a_d=6.5$\,kpc, $b_d=0.26$\,kpc, $c=0.7$\,kpc, and $d=12.0$\,kpc
(Johnston et al. 1996).
We choose these parameters for best comparison of our software with Ho00 \& Ho01.

Alternatively, we also used model number III from Bajkova \& Bobylev (2017) 
as Galactic potential,
also in Galactocentric Cartesian coordinates x,y,z:
\begin{displaymath}
\Phi _{2} = -\frac{G \cdot M_d}{\sqrt{x^2+y^2 + \Bigl(a_d + \sqrt{z^2 + b_d^2}\Bigr)^2}}
\end{displaymath}
\begin{equation} 
- \frac{G \cdot M_b}{\sqrt{x^2+y^2+z^2+b_b^2}} - \frac{G \cdot M_h}{r} \ln \Bigl(1 + \frac{r}{a_h}\Bigr)
\end{equation}
with an axisymmetric disk as first term (Miyamoto \& Nagai 1975),
a spherically symmetric bulge as second term (Hernquist 1990), 
and a massive non-truncated spherical Galactic halo as third term (Navarro et al. 1996);
for disk d, bulge b, halo h, we used the masses
M$_d=\mathrm{0.650} \cdot 10^{11}$M$_\odot$,
M$_b=\mathrm{0.103} \cdot 10^{11}$M$_\odot$,
M$_h=\mathrm{2.900} \cdot 10^{11}$M$_\odot$, respectively,
as well as the scale lengths
$a_d=\mathrm{4.40}$\,kpc,
$b_d=\mathrm{0.308}$\,kpc,
$b_b=\mathrm{0.267}$\,kpc, and
$a_h=\mathrm{7.7}$\,kpc (and gravitational constant $G$), 
derived by fitting of modern data
on rotational velocities of Galactic objects located at distances up to 200 kpc
from the Galactic center (Bajkova \& Bobylev 2017);
we choose these parameters, because they are the most recent ones incorporating Gaia data.
With any of the two potentials, 
this leads to a first-order system of six ordinary differential equations,
the integration of which is done by means of the Runge-Kutta-Fehlberg (order 4,5) method.
Alternatively, we also used the Gauss-Everhardt orbit integrator (Avdyushev 2010).
Whether we use the potential following Equ. (3) or Equ. (4) does not make significant differences
in the results -- as long as the stars remain within $\sim 1$ kpc around the Sun 
and the flight times do not exceed $\sim 10$ Myr.
Varrying the above constants within their measurement uncertainties 
(Johnston et al. 1996, Bajkova \& Bobylev 2017) also has no significant impact on our results.

A software was written from scratch to trace back the orbits of stars through the Galaxy for this work.
A sample application of it is to find possible common origins of runaway stars and
neutron stars at their birth place in supernovae,
e.g. those which contributed to the $^{60}$Fe on Earth (Neuh\"auser et al. 2012).

Our software uses C++ for both speed and interoperability with other libraries
as well as independence of licensing issues.
Third-party libraries used are the GNU Scientific Library
(GSL, www.gnu.org/software/gsl/) and Qt (www.qt.io) for convenience.
Additionally, routines for vector and matrix calculations are deployed (Steeb 1997).
 
As input values for the backward calculations of the flight paths, 
we need for each star and each OB association
the position, namely either right ascension and declination
or Galactic longitude and latitude, and then parallax
as well as the three-dimensional velocity, 
i.e. either the radial velocity (RV) with the proper motion 
(either in both equatorial or both Galactic coordinates) -- 
or, alternatively, the three-dimensional space velocities UVW. 
The RV is often not known for neutron stars -- 
unless, e.g., the inclination of a bow shock can yield a rough RV and the direction of the 3D motion
as, e.g., in the case of RXJ\,1856.5-3754 (van Kerkwijk \& Kulkarni 2001) 
and the Guitar pulsar (Tetzlaff et al. 2009),
or in cases where the neutron star is in a multiple system with another star
and the center-of-mass velocity is known as sometimes in high- and low-mass X-ray binaries 
(Liu et al. 2000, 2007).

Hence, for all neutron stars without known RV, we have to draw the RV from the Maxwell
distribution of space velocities for neutron stars
with 1D rms $\sigma = 265$ km/s (Hobbs et al. 2005)
by taking the transverse velocity into account.
If we instead take any of the other single or double-peaked distributions
(Verbunt et al. 2017),
which we also tested, we still get the same result:
no close approach between $\zeta$ Oph and PSR\,B1929+10 (Sect. 3),
but a close approach between $\zeta$ Oph and PSR\,B1706-16 (Sect. 4).
Also, when using a uniform, flat RV distribution for PSR\,B1706-16,
we can find its close approach with $\zeta$ Oph -- for a certain RV range.

For the Monte-Carlo simulation, we draw random numbers both
for the known parameters with measurement uncertainties,
namely Gaussian distributions (random numbers taken from GSL),
and also for the Maxwell distribution of neutron star space velocities
(Hobbs et al. 2005) with random numbers taken from an algorithm
implemented manually following (Mohamed 2011).

Positions, distances, and proper motions for neutron stars are listed in the
Australian Telescope National Facility (ATNF) Pulsar Catalogue
(Taylor et al. 1993, Manchester et al. 2005, 
ATNF current version 1.60);
there are currently 376 neutron stars in this catalog with both a distance
estimate and a transverse velocity, 16 of them are either too young to be
connected to the $^{60}$Fe deposition, those related to SN remnants; and
29 are too old to be related to young OB associations, namely milli-second pulsars;
hence, 331 are left for our calculations.
Most recently, Deller et al. (2019) published new astrometry for 57 pulsars,
which is not yet implemented in ATNF version 1.60; for these 44 pulsars,
we calculated backward their paths for both the previous data (as in ATNF 1.60)
and the new astrometry by Deller et al. (2019); for 17 of their pulsars,
there are no data either on parallax or proper motion in ATNF 1.60,
so that we added those 17 objects to the above list of 331 neutron stars (then 348).
Pulsar distances are either independent estimates like a parallax
with measurement uncertainties or, otherwise, from the dispersion measure 
(latest in Stovall et al. 2015) --
for those distances, we use $10\%$ error bars following
tables 6 \& 7, equation 47, and figure 14 from Yao et al. (2017).

Then, there are 68 pulsars, namely 50 in addition to those mentioned above,
for which there is a Gaia DR2 
entry very close to the pulsar position --
as listed on Simbad as cross-identifications between pulsars and Gaia stars.
The matches between the pulsar positions (ATNF) and Gaia DR2 stars were
selected as follows: high-quality astrometric data with precision up to $1^{\prime \prime}$,
and the positional offset between the pulsar and the Gaia star must be smaller than $1^{\prime \prime}$.
These 68 pulsars are listed in Table 1.
In most of these cases, Gaia has detected the pulsar in the optical,
in other cases the Gaia star is a companion to the neutron star forming a low-mass
or high-mass X-ray binary.

Then, we also added PSR\,J0002+6216 (Schinzel et al. 2019)
and the faint object, possibly a planet candidate, close to
Fomalhaut, which can be a background neutron star (Neuh\"auser et al. 2015).
In total, we then have 400 neutron stars
with sufficient astrometric data.

Positions, distances, and 3D space motions of the three
sub-groups of Scorpius-Centaurus-Lupus are found in Wright \& Mamajek (2018)
based on Gaia data.

\onecolumn

\begin{longtable}{llrrrcc}
\multicolumn{7}{l}{{\bf Table 1.} Data of 68 additional neutron stars: pulsars within 1 arc sec of Gaia stars} \\
\multicolumn{7}{l}{(we list the distances, proper motions, and G magnitudes of the Gaia stars)} \\ \hline
%
Pulsar & Gaia DR2 source & distance & \multicolumn{2}{c}{proper motion} & G & Simbad    \\ 
       &                 & [pc] (a) & $\mu _{\alpha}^{*}$ [mas/yr] & $\mu _{\delta}$ [mas/yr] & [mag] & class \\ \hline
\endfirsthead
\multicolumn{7}{l}{Data of 68 additional neutron stars: pulsars at the positions of Gaia stars (table 1 continued)} \\ \hline
Pulsar & Gaia DR2 source & distance & \multicolumn{2}{c}{proper motion}                   & G     & Simbad  \\ 
       &                 & [pc] (a) & $\mu _{\alpha}^{*}$ [mas/yr] & $\mu _{\delta}$ [mas/yr] & [mag] & class \\ \hline
\endhead
\hline
\endfoot
\endlastfoot
B0329+54  * &  445135392719883904 & 1163.57$_{-703}^{+1854}$  &  $ -0.2 \pm  2.8 $ & $    -3.8 \pm  2.0$& 20.1861 & psr  \\ 
J0348+0432 *+ & 3273288485744249344 & 1402.58$_{-485}^{+726}$   &  $  3.1 \pm  2.1 $ & $    -0.1 \pm  1.5$& 20.6387 & psr \\  
J1001-5559  & 5259965255350026368 & 3203.94$_{-1130}^{+2128}$ &  $ -5.7 \pm  0.3 $ & $     4.64 \pm  0.37$& 18.6803 & psr \\ 
J1012+5307 *+ &  851610861391010944 & 734.40$_{-185}^{+330}$    &  $  3.0 \pm  0.5 $ & $   -26.94 \pm  0.63$& 19.6256 & psr           \\ 
      J1012-5830  & 5258214317448912000 & 2894.91$_{-1677}^{+2833}$ &  $ -3.1 \pm  1.9 $ & $     3.19 \pm  1.93$& 20.5227 & psr       \\ 
      J1023+0038 *+ & 3831382647922429952 & 1266.40$_{-196}^{+273}$   &  $  4.8 \pm  0.1 $ & $   -17.35 \pm  0.14$& 16.2651 & LMXB    \\ 
      B1030-58    & 5255318414303381888 & 1315.46$_{-51}^{+55}$     &  $ -7.0 \pm  0.1 $ & $     3.226 \pm  0.053$& 11.6982 & psr              \\ 
      J1048+2339 *+ & 3990037124929068032 & 851.84$_{-315}^{+592}$    &  $-16.2 \pm  1.0 $ & $   -11.7 \pm  1.3$& 19.6526   & psr                \\ 
      J1054-5943  & 5338371090272184320 & 1414.95$_{-42}^{+44}$     &  $ -3.9 \pm  0.0 $ & $    -3.059 \pm  0.033$& 14.3615 & psr                 \\ 
       J1130-6807 & 5235043660391986944 & 2795.03$_{-592}^{+985}$   &  $ -7.3 \pm  0.1 $ & $     0.49 \pm  0.13$& 17.1798   & psr           \\ 
       J1144-6146 & 5334905773233011328 & 628.73$_{-13}^{+14}$      &  $-15.1 \pm  0.04$ & $    -5.554 \pm  0.043$& 14.7615 & psr             \\ 
       J1156-5909 & 5341440995453204096 & 2138.59$_{-370}^{+552}$   &  $ -4.4 \pm  0.1 $ & $     1.04 \pm  0.13$& 17.3685   & psr               \\ 
       J1227-4853 *+ & 6128369984328414336 & 1577.57$_{-372}^{+658}$   &  $-18.7 \pm  0.2 $ & $     7.39 \pm  0.12$& 18.0758 & LMXB                 \\ 
       J1237-6725 & 5856816764375829376 & 1770.04$_{-155}^{+188}$   &  $ -0.69 \pm  0.09 $ & $     5.385 \pm  0.076$& 16.2430 & psr               \\ 
       B1240-64   & 5862634658360519168 & 1663.08$_{-100}^{+114}$   &  $ -1.990 \pm  0.054 $ & $    -1.341 \pm  0.049$& 15.6859 & psr              \\ 
       B1259-63 *+ & 5862299960127967488 & 2243.23$_{-148}^{+170}$   &  $ -6.986 \pm  0.043 $ & $    -0.416 \pm  0.044$&  9.6251 & HMXB              \\ 
       J1335-3642 & 6164329241532641536 (b) & 406.50$_{-10}^{+11}$  &  $-25.86 \pm  0.14 $ & $    -9.27 \pm  0.11$& 10.9616 & psr        \\ 
       J1347-5947 & 5870519526774500736 & 3250.99$_{-1159}^{+2094}$ &  $ -8.29 \pm  0.20 $ & $    -1.79 \pm  0.35$& 18.5835 & psr               \\ 
       J1355-6206 & 5866295791532128640 & 2827.15$_{-989}^{+1966}$  &  $ -4.13 \pm  0.18 $ & $    -2.49 \pm  0.26$& 18.3315 & psr              \\ 
       B1359-51   & 6090172744223852800 (c) & 545.53$_{-13}^{+14}$  &  $-21.659 \pm  0.082 $ & $   -11.785 \pm  0.083$& 10.9330 & psr         \\ 
       J1425-6210 & 5854330424971261696 & 3273.62$_{-619}^{+944}$   &  $ -0.994 \pm  0.082 $ & $    -4.32 \pm  0.13$& 16.6219  & psr                \\ 
       J1444-5941 & 5878936116484167040 & 5657.40$_{-1638}^{+2466}$ &  $ -6.42 \pm  0.24 $ & $    -6.66 \pm  0.21$& 17.5435    & psr              \\ 
       B1451-68 * & 5800163843613786496 & 4199.41$_{-833}^{+1289}$  &  $ -8.090 \pm  0.067 $ & $    -4.975 \pm  0.082$& 16.2627 & psr               \\ 
       J1509-6015 & 5876497399692841088 & 5268.84$_{-1733}^{+2744}$ &  $ -5.23 \pm  0.16 $ & $    -2.35 \pm  0.22$& 17.7991   & psr             \\ 
       B1530-53 * & 5885592598562423552 & 2869.62$_{-658}^{+1111}$  &  $ -5.87 \pm  0.23 $ & $    -4.13 \pm  0.22$& 15.1179   & psr            \\ 
       J1538-5519 & 5885107095440413056 & 1729.63$_{-328}^{+515}$   &  $ -9.36 \pm  0.29 $ & $    -7.41 \pm  0.19$& 17.3774   & psr           \\ 
       J1548-5607 & 5884168520525984256 & 2641.68$_{-1673}^{+2790}$ &  $  2.1 \pm  2.1 $ & $    -8.06 \pm  2.32$& 20.5368     & psr            \\ 
       J1616-5017 & 5935105286461295744 & 3090.64$_{-1102}^{+2100}$ &  $ -2.61 \pm  0.31 $ & $    -3.32 \pm  0.20$& 18.3809   & psr             \\ 
       B1626-47   & 5941975447438391424 & 3475.18$_{-1597}^{+2545}$ &  $-12.3 \pm  2.9 $ & $    -3.6 \pm  1.4$& 20.2151       & psr         \\ 
       J1632-4757 & 5941203422793565696 & 2379.04$_{-582}^{+1040}$  &  $ -1.08 \pm  0.17 $ & $    -2.42 \pm  0.11$& 17.1185   & psr             \\ 
       J1650-4126 & 5968092495059303680 & 870.83$_{-32}^{+34}$      &  $ -7.989 \pm  0.068 $ & $   -10.983 \pm  0.048$& 14.7391 & psr                  \\ 
       B1657-45   & 5963300303657586560 & 1314.67$_{-907}^{+2935}$  &  $ -2.1 \pm  2.8 $ & $     0.9 \pm  1.6$& 20.3448       & psr        \\ 
       J1707-4341 & 5965148788863345664 & 2986.09$_{-1507}^{+2639}$ &  $ -4.7 \pm  1.2 $ & $    -4.02 \pm  0.83$& 20.0103     & psr           \\ 
       J1713-3844 & 5973093589309766144 & 1582.50$_{-80}^{+89}$     &  $ -0.279 \pm  0.050 $ & $    -2.594 \pm  0.034$& 13.7786 & psr                  \\ 
       J1720-3659 & 5975156612410606976 & 2342.23$_{-915}^{+2260}$  &  $ -0.26 \pm  0.41 $ & $    -0.73 \pm  0.28$& 18.1837     & psr              \\ 
       J1723-2837 *+ & 4059795674516044800 & 907.89$_{-44}^{+49}$      &  $-11.713 \pm  0.082 $ & $   -23.99 \pm  0.06$& 15.5491 & psr                  \\ 
       J1725-3848 & 5972519128840395136 & 3773.07$_{-1617}^{+2841}$ &  $ -1.73 \pm  0.70 $ & $    -3.69 \pm  0.46$& 18.8589    & psr            \\ 
       J1736-2457 & 4062231440714447232 & 6348.10$_{-2503}^{+4103}$ &  $ -1.85 \pm  0.59 $ & $    -7.95 \pm  0.50$& 17.0707    & psr            \\ 
       J1738-2736 & 4060535272241504256 & 4934.67$_{-2489}^{+4259}$ &  $ -3.18 \pm  0.77 $ & $    -8.94 \pm  0.64$& 18.9744    & psr           \\ 
       J1743-4212 & 5957265359205874688 & 4825.17$_{-2026}^{+3671}$ &  $ -3.28 \pm  0.37 $ & $    -0.19 \pm  0.29$& 18.0058    & psr           \\ 
       J1744-3130 & 4055849733443329024 & 1360.40$_{-146}^{+186}$   &  $  1.18 \pm  0.16 $ & $    -2.81 \pm  0.13$& 16.3199    & psr           \\ 
       J1755-3716 + & 4037273763727266816 & 5767.54$_{-2613}^{+4207}$ &  $ -2.92 \pm  0.97 $ & $    -5.71 \pm  0.80$& 19.0385  & psr             \\ 
       J1806-1618 & 4144356335016612608 & 5515.17$_{-1768}^{+2797}$ &  $ -4.48 \pm  0.25 $ & $    -6.25 \pm  0.21$& 17.1189    & psr           \\ 
       J1808-1020 & 4157239342774281344 & 3286.27$_{-1379}^{+2517}$ &  $  3.60 \pm  0.57 $ & $    -3.57 \pm  0.47$& 18.8174    & psr            \\ 
       J1809-0743 & 4159031649807119488 & 3957.14$_{-1560}^{+2509}$ &  $ -0.87 \pm  0.51 $ & $    -6.27 \pm  0.47$& 18.8724    & psr          \\ 
       J1810+1744 + & 4526229058440076288 & 1635.50$_{-879}^{+2070}$  &  $  6.4 \pm  1.8 $ & $    -7.2 \pm  2.1$& 20.0831      & psr         \\ 
       J1818-1519 & 4146162862638980864 & 4190.71$_{-1688}^{+2733}$ &  $ -0.35 \pm  0.57 $ & $    -2.73 \pm  0.51$& 18.6578    & psr           \\ 
       J1824-1423 & 4104296586587171200 & 2570.13$_{-1162}^{+2561}$ &  $ -1.10 \pm  0.76 $ & $    -4.83 \pm  0.71$& 18.9830    & psr           \\ 
       J1839-0459 & 4256598742584390272 & 2468.85$_{-804}^{+1711}$  &  $ -3.76 \pm  0.24$ & $    -3.50 \pm  0.21$& 17.4634     & psr          \\ 
       J1839-0905 & 4155609699080401920 & 1718.22$_{-203}^{+264}$   &  $ -3.46 \pm  0.12 $ & $    -4.64 \pm  0.11$& 16.5294    & psr            \\ 
       J1841-0524 & 4256511189624452608 & 2609.49$_{-1396}^{+2685}$ &  $  7.39 \pm  0.93 $ & $   -23.01 \pm  0.84$& 13.5645    & psr            \\ 
       B1839-04   & 4258439045825710208 & 3331.39$_{-747}^{+1229}$  &  $ -0.74 \pm  0.13 $ & $    -3.616 \pm  0.095$& 15.0533  & psr             \\ 
       B1844-04 * & 4258299824471461760 & 1943.72$_{-464}^{+840}$   &  $ -1.05 \pm  0.20 $ & $     0.88 \pm  0.16$& 17.0514    & psr           \\ 
       J1900+0227 & 4268630247305381760 & 361.85$_{-24}^{+28}$      &  $ 36.92 \pm  0.36 $ & $  -108.38 \pm  0.30$& 18.0218    & psr           \\ 
       J1901-1740 & 4088222589900106496 & 2452.70$_{-780}^{+1811}$  &  $ -0.24 \pm  0.20 $ & $    -2.94 \pm  0.16$& 17.3085    & psr           \\ 
       B1902-01 * & 4262491708247480320 & 610.48$_{-19}^{+20}$      &  $ -2.03 \pm  0.13 $ & $   -12.279 \pm  0.094$& 11.2649  & psr             \\ 
       B1907+10 * & 4312461286246658176 & 3390.87$_{-936}^{+1665}$  &  $ -2.93 \pm  0.20 $ & $     0.42 \pm  0.20$& 17.3128    & psr           \\ 
       B1907-03   & 4260681229236436608 & 400.42$_{-12}^{+13}$      &  $  2.17 \pm  0.15 $ & $   -42.74 \pm  0.12$& 16.7506    & psr           \\ 
       B1919+20   & 4515936422712632832 & 3823.84$_{-1667}^{+2733}$ &  $ -2.80 \pm  0.48 $ & $    -5.25 \pm  0.87$& 19.5474    & psr           \\ 
       B1930+22   & 2018583673636873472 & 2072.52$_{-759}^{+1959}$  &  $ -6.31 \pm  0.33$ & $    -3.62 \pm  0.34$& 18.7900    & psr           \\ 
       B1933+16 * & 4321800842917139456 & 3792.91$_{-1338}^{+2305}$ &  $ -3.55 \pm  0.27 $ & $    -4.99 \pm  0.30$& 18.4111   & psr            \\ 
       J1958+2846 & 2030000280820200960 & 2018.88$_{-880}^{+2248}$  &  $ -4.03 \pm  0.53 $ & $    -6.12 \pm  0.65$& 19.3597   & psr            \\ 
       B2016+28 * & 1837105534255815296 & 4295.66$_{-1420}^{+2152}$ &  $ -3.77 \pm  0.42 $ & $    -5.05 \pm  0.49$& 19.0070   & psr           \\ 
       J2027+4557 & 2071054503122390144 & 1840.53$_{-114}^{+130}$   &  $  2.126 \pm  0.076 $ & $    -2.059 \pm  0.061$& 15.7254    & psr                \\ 
       J2032+4127 *+ & 2067835682818358400 & 1385.26$_{-62}^{+68}$     &  $ -2.991 \pm  0.048 $ & $    -0.742 \pm  0.055$& 11.3611 & Be star (e)              \\ 
       J2033+1734 *+ & 1812057151844384512 (d) & 906.20$_{-26}^{+27}$  &  $  5.187 \pm  0.052 $ & $     1.518 \pm  0.053$& 10.3242 & psr     \\ 
       J2215+5135 + & 2001168543319218048 & 2471.98$_{-1009}^{+1863}$ &  $  0.31 \pm  0.54 $ & $     1.88 \pm  0.60$& 19.2408      & psr         \\ 
       J2339-0533 *+ & 2440660623886405504 & 1183.68$_{-280}^{+460}$   &  $  4.15 \pm  0.48 $ & $   -10.31 \pm  0.31$& 18.9704     & psr          \\ \hline
       


\multicolumn{7}{l}{Notes: Pulsars sorted by right ascension, $\mu _{\alpha}^{*}$ is $\mu _{\alpha} \cdot \cos \delta$ (in total, we considered 400 neutron stars).} \\
\multicolumn{7}{l}{(a) Gaia DR2 distances (Bailer-Jones et al. 2018).} \\
\multicolumn{7}{l}{Proper motions and Gaia G mag from Gaia DR2 (Gaia Collaboration: Brown et al. 2018).} \\
\multicolumn{7}{l}{Simbad: pulsar (psr), LMXB (low-mass X-ray binary), or HMXB (high-mass X-ray binary).} \\
\multicolumn{7}{l}{(e) Be star [MT91] 213 in binary with pulsar PSR J2032+4127.} \\
\multicolumn{7}{l}{Radial velocity according to Gaia: (b) $-44.57 \pm 0.55$ km/s, (c) $-9.57 \pm 0.56$ km/s, (d) $-57.71 \pm 0.47$ km/s.} \\
\multicolumn{7}{l}{(*) These 18 pulsars also have parallax and/or proper motion in the pulsar catalog (Manchester et al. 2005), we calculated} \\
\multicolumn{7}{l}{backwards their motion separately both for the astrometric data of the pulsar and for the astrometric data of the Gaia star.} \\
\multicolumn{7}{l}{(+) These 13 pulsars are already marked as binaries in the pulsar catalog (Manchester et al. 2005).} 

\end{longtable}

\twocolumn

The type of coordinates used for input determines which parameters are varied during the Monte-Carlo simulation.
Positional uncertainties of the objects are negligible (Ho00, Ho01).
The error-prone observables RV and proper motions or space velocities
are varied with a Gaussian distribution according to their uncertainties,
the parallax also as truncated Gaussian, restricted liberally to 0.1-25 kpc;
for the RVs of neutron stars, we use a Maxwellian distribution
of their space velocity, see above.
For both Hipparcos and Gaia data,
we take into account correlations between the astrometric parameters
as given in the respective covariance matrix.

After the variation the coordinates are converted for the calculation into the right-handed
Galactic Cartesian system X, Y, Z at present with the corresponding velocities U, V, W (Johnson \& Soderblom 1987).
As distance of the Local Standard of Rest (LSR) to the Galactic center, we use 8.5 kpc (Verbiest et al. 2012),
and as velocity around it 220 km/s (Binney \& Tremaine 1987).
The coordinates of the Sun relative to the LSR are (X, Y, Z) = (0, 0, 15) pc (Voigt 2012),
its velocity in the LSR (U, V, W) = (10.0, 5.25, 7.17) km/s (Dehnen \& Binney 1998).
Observed coordinates and velocities are converted accordingly.
Just as the constants in the Galactic potential, varrying the above Galactic constants 
within their measurement uncertainties does not have any significant impact on the results.

Monte-Carlo simulations were done for either two (runaway and neutron star)
or three objects (runaway star, neutron star, and an OB association), 
all being traced backwards in time.

The `success' of a simulation run can be rated according to certain criteria such as
the number of close encounters within, e.g., 10 pc (Ho0, Ho01) or 15 pc (Tetzlaff et al. 2010).

Given the measurement uncertainties of the three-dimensional space motion and
parallax of all objects involved,
and given that the number of trails in any Monte Carlo
simulation is finite, one cannot expect a separation between two objects
(i.e. between two three-dimensional Gauss curves) of 0.0 pc,
but a separation of less than a few pc can indicate a close encounter,
i.e. a common origin in a binary SN (Ho0, Ho01, Tetzlaff et al. 2010).
Given the variation of the input parameters, even for a large number of runs, here 3 million for each case,
the number of runs with a small minimum separation is expected to be small,
e.g. only $0.14\%$ of simulations in Ho00 and Ho01 for $\zeta$ Oph and PSR\,B1929+10
yielded encounters within 10 pc, but were consistent with
the hypothesis that pulsar and runaway star were at the same time at the same position.

We also compute the likelihood for each run:
in order to determine most likely past flight paths of runaway stars,
neutron stars, and associations among the many simulations,
we used as measure the multivariate Gaussian likelihood.
This likelihood is the sum of likelihoods of the input parameters and their
uncertainties with covariance matrices, as well as output parameters,
i.e. positions and the time of a close approach within a
stellar association or subgroup.
We use these likelihoods to compare the probability of different runs with eachother.

For most neutron stars, the RV is not known, 
so that any solution is valid only for a certain RV range.
Before we can conclude with some certainty that both a certain neutron star and 
a certain runaway star were ejected in a SN at a certain 
time and position, the following tests have to be passed: 
\begin{enumerate}
\item For both the runaway star as well as the neutron star,
the 3D coordinates in space and the 3D space velocity
components must be know --
except the RV of the neutron star, which is usually unknown,
\item The runaway star should to be outside of an OB association at the present time 
-- depending on the time passed since the SN and the space velocity.
\item Both a neutron star and a runaway star have to be at the same time 
(within the given measurement uncertainties or error bars) at the same 
three-dimensional region (within a few pc) -- for a sufficiently large number of runs.
\item The close approach between the neutron star and the runaway star
has to be in a three-dimensional volume in space, where an OB association
has been located at the time of the close approach.
\item The close approach should be located inside a cavity of the interstellar
medium, a low-density bubble, or even a SN remnant.
\item The age of the runaway star should be consistent with the age
of the OB association.
\item There must be no other possible solution for the origin of the neutron star
with any other runaway star with a nearly equal or even larger percentage of runs.
\item There must be no other possible solution for the origin of the runaway star
with any other neutron star with a nearly equal or even larger percentage of runs.
\item If the close approach (supernova) is inside a SN remnant, its age must be
consistent with the flight time (kinematic age) -- 
or, otherwise, if the close approach (supernova) is outside a SN remnant, 
the flight time should be larger than $\sim 100,000$ yr, the maximal age of SN remnants --
depending on the neutron star velocity.
\item The difference between the flight time, i.e the kinematic age of the neutron star,
and the association age must be at least $\sim 3$ Myr,
i.e. the life-time of most massive stars until their core-collapse SN, and at most
$\sim 30$ Myr, i.e. the life-time of the lowest-mass (electron-capture) SN progenitors.
\item The flight time must not be significantly larger than the characteristic
spin-down age, i.e. the upper age limit, of the neutron star; for $\sim 1$ to few
Myr old neutron stars, the characteristic age is normally comparable to the true age.
\item The mass of the SN progenitor star from the difference between
association age and flight time as life-time of the progenitor must not
be smaller than the mass of the runaway star.
\item The mass of the SN progenitor should not be unexpectedly large given
the present mass function of the host association compared to the IMF.
\item The sum of the mass of the progenitor star and the runaway star has
to be at least twice the sum of the masses of the runaway star and the
neutron star ($\sim 1.4$ M$_{\odot}$, the typical neutron star mass), 
so that the progenitor binary system got unbound due to the SN
even without an additional kick.
\item The space velocity range of the neutron star, for which the close approach
is possible, should be 
consistent with the typical mean velocity of neutron stars
within the measurement uncertainties.
\item The space velocity of the runaway star should not point back
to the center of the Galaxy,
where it could have been ejected by the massive Black Hole.
\end{enumerate}
The solution presented here, a common origin of $\zeta$ Oph and 
PSR\,B1706-16 (Sect. 4), indeed passes all 16 tests -- and $\zeta$ Oph is a single
star (typical for runaways) with large rotational velocity and He overabundance 
(Marcolino et al. 2009),
possibly indicating previous mass exchange from a nearby SN progenitor.
Multiplicity alone would otherwise not be a reason to reject a star as a runaway
ejected from the SN in a multiple star, because close multiples can also get ejected;
in case of spectroscopic binaries, we use the center-of-mass velocity if known;
e.g. V716 Cen was previously thought to originate from 
Scorpius-Centaurus-Lupus (Ho00, Ho01)
using RV=$66$ km/s, but the orbit of this spectroscopic binary was later solved (Bakis et al. 2008), 
who found a center-of-mass velocity of $-10.3 \pm 6.9$ km/s (which we used).
We calculated backward the flight path for all those ten (of a total of 56) runaway stars (Ho00, Ho01),
which come from or at least near the Sco-Cen-Lup association (see Sect. 4),
and all 400 Galactic neutron stars with known transverse velocity (which are neither too young
nor too old) -- all tested against Sco-Cen-Lup. 
None of the other close meetings of runaway stars and neutron stars passed all 16 tests;
there were not even any other pairs close to passing most tests.

\section{Testing our software with $\zeta$ Oph and PSR\,B1929+10}

We first test our software for backward calculation of flight paths 
with the input values of the original case (Ho00, Ho01), see our Table 2. 

As in the original case (Ho00, Ho01),
we increased the error bars on the proper motion of the pulsar PSR\,B1929+10 
by a factor of two 
(just because this was also done that way in Ho00 \& Ho01)
and used as RV of the pulsar of $200 \pm 50$ km/s as well as a parallax 
of $4 \pm 2$ mas
(this value was also used by Ho00 \& Ho01, which we use to validate our software),
the latter being consistent with the published value of 
$5 \pm 1.5$ mas (Campbell 1995, Campbell et al. 1996). 
Only for these parameters, $\zeta$ Oph and PSR\,B1929+10 would have been ejected 
from the same location at the same time $\sim 1$ Myr ago (Ho00, Ho01).
We show in case 1 of Table 2 and in Fig. 1 
that our software can confirm the results from previous calculations (Ho00, Ho01)
when using the same input values: 
$\zeta$ Oph and PSR\,B1929+10 would have been at the same position within 
UpSco $\sim 1$ Myr ago.

We also calculated backward the motions of $\zeta$ Oph, the neutron star PSR\,B1929+10, 
and UpSco for all sets of input values published so far (Table 2),
including Gaia data for $\zeta$ Oph (Gaia 2018) and UpSco (Wright \& Mamajek 2018)
See Table 2 for the input values and results for the different cases.

\begin{table*}
\caption{Calculating backward the motion of $\zeta$ Oph and PSR\,B1929+10}
\begin{tabular}{lccccccc} \hline
Object & \multicolumn{2}{c}{Coordinates} & Parallax & \multicolumn{2}{c}{Proper motion or U,V [km/s]} & Rad. Vel. & \hspace{-1.0cm} Rem. \\
name   & $\alpha$ & $\delta$             & $\pi$     & $\mu _{\alpha} \cdot \cos \delta$ & $\mu _{\delta}$  & or W & \\ 
       & [hh:mm:ss] & [$^{\circ}$:$^{\prime}$:$^{\prime \prime}$]  & [mas]     & [mas/yr]                 & [mas/yr] & [km/s]    & \\ \hline
\multicolumn{8}{l}{Case 1: original case (Ho00, Ho01), Fig. 1:} \\
$\zeta$ Oph & 16:37:9.53  & $-10$:34:1.7 & $7.12 \pm 0.71$ & $13.07 \pm 0.85$ mas/yr & $25.44 \pm 0.72$ mas/yr & $-9.0 \pm 5.5$ km/s & \hspace{-0.9cm} a \\
B1929+10 & 19:32:13.87 & 10:59:31.8 & $4 \pm 2$       & $99 \pm 12$ mas/yr & $39 \pm 8$ mas/yr & $200 \pm 50$ km/s & \hspace{-0.9cm} b \\ 
UpSco       & $l = 351.07^{\circ}$ & $b = 19.43^{\circ}$ & $6.9 \pm 0.12$ & $\mu _{l}^*$=-$24.5 \pm 0.1$ mas/yr & $\mu _{b}$=-$8.1 \pm 0.1$ mas/yr & $-4.6$ km/s & \hspace{-0.9cm} c \\
\multicolumn{8}{l}{\small{Ho00, Ho01: for 4214 of 3 million runs, both stars within 10\,pc of UpSco center (min. 0.35\,pc, $\sim\,1$\,Myr ago)}} \\
\multicolumn{8}{l}{\small{Our result: for $\sim 4100$ of 3 million runs, both stars within 10\,pc of UpSco center (min. 0.09\,pc, 1.12\,Myr ago)}} \\ \hline
\multicolumn{8}{l}{Case 2: as in Tetzlaff et al. (2010) with new astrometry for pulsar, and a Maxwell-distributed RV:} \\
$\zeta$ Oph & 16:37:9.53  & $-10$:34:1.7 & $7.12 \pm 0.71$ & $13.07 \pm 0.85$ mas/yr & $25.44 \pm 0.72$ mas/yr & $-9.0 \pm 5.5$ km/s & \hspace{-0.7cm} a \\
B1929+10 & 19:32:13.95 & 10:59:32.4 & $2.76 \pm 0.14$ & $94.03 \pm 0.14$ mas/yr & $43.37 \pm 0.29$ mas/yr & -- & \hspace{-0.9cm} d,e  \\
UpSco       & $l = 351.07^{\circ}$ & $b = 19.43^{\circ}$ & 6.65 & U=-$6.7 \pm 5.9$ km/s & V=-$16.0 \pm 3.5$ km/s & W=-$8 \pm 2.7$ km/s & \hspace{-0.7cm} f \\ 
\multicolumn{8}{l}{\small{Tetzlaff  et al. (2010): for 3 million runs, closest approach between the two stars still 4\,pc, too large to have met}} \\
\multicolumn{8}{l}{\small{Our result: in $\sim 700$ of 3 million runs, both stars within 15\,pc from UpSco center, closest approach 4\,pc, 0.42\,Myr ago}} \\ \hline
\multicolumn{8}{l}{Case 3: with new Hipparcos astrometry for $\zeta$ Oph, pulsar error bars multiplied by 10:} \\
$\zeta$ Oph & 16:37:9.539 & $-10$:34:1.5 & $8.91 \pm 0.2$ & $15.26 \pm 0.26$ mas/yr & $24.79 \pm 0.22$ mas/yr & $-9.0 \pm 5.5$ km/s & \hspace{-0.7cm} g \\
B1929+10 & 19:32:13.95 & 10:59:32.4 & $2.76 \pm 1.4$ & $94.03 \pm 1.4$ mas/yr & $43.37 \pm 2.9$ mas/yr & -- & \hspace{-0.9cm} h,e  \\
UpSco      & $l = 351.07^{\circ}$ & $b = 19.43^{\circ}$ & 6.65 & U=-$6.7 \pm 5.9$ km/s & V=-$16.0 \pm 3.5$ km/s & W=-$8 \pm 2.7$ km/s & \hspace{-0.7cm} f \\
\multicolumn{8}{l}{\small{Our result: for $\sim 60$ of 3 million runs, both stars within 15 pc from UpSco center, closest 0.3\,pc, 0.64\,Myr ago}} \\ \hline
\multicolumn{8}{l}{Case 4: with new Hipparcos astrometry and new RV for $\zeta$ Oph, pulsar error bars multiplied by 10:} \\
$\zeta$ Oph & 16:37:9.54 & $-10$:34:1.5 & $8.91 \pm 0.2$ & $15.26 \pm 0.26$ mas/yr & $24.79 \pm 0.22$ mas/yr & $12.2 \pm 3.3$ km/s & \hspace{-0.8cm} g,i \\
B1929+10 & 19:32:14.02 & 10:59:32.9 & $2.77 \pm 0.8$ & $94.08 \pm 1.7$ mas/yr & $43.25 \pm 1.6$ mas/yr & -- & \hspace{-0.8cm} e,j \\
UpSco      & $l = 351.07^{\circ}$ & $b = 19.43^{\circ}$ & 6.65 & U=-$6.7 \pm 5.9$ km/s & V=-$16.0 \pm 3.5$ km/s & W=-$8 \pm 2.7$ km/s & \hspace{-0.7cm} f \\
\multicolumn{8}{l}{\small{Our result: for no orbits at all, the stars get within 15 pc of UpSco center}} \\ \hline
\multicolumn{8}{l}{Case 5: with new RV and new Gaia DR2 astrometry for $\zeta$ Oph, pulsar error bars multiplied by 10:} \\
$\zeta$ Oph & 16:37:9.54 & $-10$:34:1.5 & $5.8 \pm 1.0$ & $27.7 \pm 1.7$ mas/yr & $37.2 \pm 1.2$ mas/yr & $12.2 \pm 3.3$ km/s & \hspace{-0.7cm} i,k \\
B1929+10 & 19:32:14.02 & 10:59:32.9 & $2.77 \pm 0.8$ & $94.08 \pm 1.7$ mas/yr & $43.25 \pm 1.6$ mas/yr & -- & \hspace{-0.7cm} e,j \\
UpSco    & $l = 351.07^{\circ}$ & $b = 19.43^{\circ}$ & $6.99 \pm 0.02$ & U=-$6.16 \pm 1.63$ km/s & V=-$16.89 \pm 1.14$ km/s & W=-$7.05 \pm 2.51$ km/s & \hspace{-0.5cm} l \\
\multicolumn{8}{l}{\small{Our result: for $\sim$ 2000 of 3 million runs, both stars within 15 pc to the center of UpSco simultaneously;}} \\
\multicolumn{8}{l}{\small{minimum separation 2.5 pc 0.49 Myr ago -- too far for a binary SN}} \\ \hline 
\multicolumn{8}{l}{Case 6: with new RV and new Gaia DR2 astrometry for $\zeta$ Oph, pulsar error bars as published (not multiplied):} \\
$\zeta$ Oph & 16:37:9.54 & $-10$:34:1.5 & $5.8 \pm 1.0$ & $27.7 \pm 1.7$ mas/yr & $37.2 \pm 1.2$ mas/yr & $12.2 \pm 3.3$ km/s & \hspace{-0.7cm} i,k \\
B1929+10 & 19:32:14.02 & 10:59:32.9 & $2.77 \pm 0.08$ & $94.08 \pm 0.17$ mas/yr & $43.25 \pm 0.16$ mas/yr & -- & \hspace{-0.7cm} e,j \\
UpSco    & $l = 351.07^{\circ}$ & $b = 19.43^{\circ}$ & $6.99 \pm 0.02$ & U=-$6.16 \pm 1.63$ km/s & V=-$16.89 \pm 1.14$ km/s & W=-$7.05 \pm 2.51$ km/s & \hspace{-0.5cm} l \\
\multicolumn{8}{l}{\small{Our result: for $\sim$1000 of 3 million runs, both stars within 15 pc to the center of UpSco simultaneously;}} \\
\multicolumn{8}{l}{\small{minimum separation 5.1 pc 0.42 Myr ago -- too far for a binary SN.}} \\ \hline
\end{tabular}

\smallskip

Remarks: 
(a) Astrometry from the Hipparcos catalog (ESA 1997),
    RV from the Hipparcos input catalog (Turon et al. 1992). 
(b) Positions and proper motions from Taylor et al. (1993), 
parallax and RV assumed, proper motion measurement uncertainties as given here 
and as used in the calculations were multiplied by 2 as in Ho00 and Ho01. 
(c) from de Zeeuw et al. (1999), $\mu _{l}^*$ is $\mu _{l} \cdot \cos b$. 
(d) New astrometry from Chatterjee et al. (2004),
    here the more conservative data from the 2nd line in their table 1. 
(e) RV Maxwell-distributed (Hobbs et al. 2005, Tetzlaff et al. 2010).
(f) $UVW$ [km/s] for UpSco from Sartori et al. (2003) as given in Fernandez et al. (2008). 
(g) New Hipparcos astrometry (van Leeuwen 2007).
(h) Same as in case 2, but error bars for parallax and proper motion multiplied by 10 as in 
    Tetzlaff et al. (2010), so that the two stars can meet. 
(i) New RV for $\zeta$ Oph from Zehe et al. (2018). 
(j) Newest astrometry for pulsar (Kirsten et al. 2015). 
(k) New Gaia DR2 astrometry for $\zeta$ Oph (Gaia 2018),
which is, however, not reliable due to large brightness of $\zeta$ Oph (Lindegren 2018). 
(l) New Gaia astrometry for UpSco with a radius of 17.1 pc (Wright \& Mamajek 2018),
which is well reliable. 
(k,l) The systematic offset of $-0.029$ mas to the parallax zero-point in 
Gaia DR2 data (Lindegren et al. 2018) 
would be only 1 pc here, i.e. well within the error bars.
\end{table*}

In the original calculations (Ho00, Ho01), 
a RV of $\zeta$ Oph of $-9.0 \pm 5.5$ km/s was used. 
The new value for the RV of the star, however, is $+12.2 \pm 3.3$ km/s (Zehe et al. 2018), 
obtained with 48 high-resolution spectroscopic observations from 2015 July to 2016 Sep, 
where the RV was constant (Zehe et al. 2018), 
also rejecting the claim that $\zeta$ Oph would be 
a spectroscopic binary (Chini et al. 2012, 
based on one spectrum only: 
some line shapes are variable due to strong stellar wind --
the spectral type of $\zeta$ Oph was classified 
from O9.5Vnn (Morgan et al. 1943, Marcolio et al. 2009)
to O9.2IVnn star (Sota et al. 2014),
where it was also found not to be a spectroscopic binary. 
The previously published RV values for $\zeta$ Oph can be grouped into two bins, 
one peaking between $-20$ to $-5$ km/s and one peaking between $+10$ and $+20$ km/s
(see Zehe et al. 2018). 
While the latter range includes the new RV value (Zehe et al. 2018) 
of the star, the former includes the RV of the interstellar Na lines. 
In the original work (Ho00, Ho01), 
an RV of $-9.0 \pm 5.5$ km/s was used, i.e. strongly different from the new value (Zehe et al. 2018).  
When using the new RV of $\zeta$ Oph (Zehe et al. 2018), 
see cases 3 to 6 in Table 2, PSR\,B1929+10 and $\zeta$ Oph cannot have been 
at the same time at the same position.

\begin{figure}
\centering
\includegraphics[width=\columnwidth]{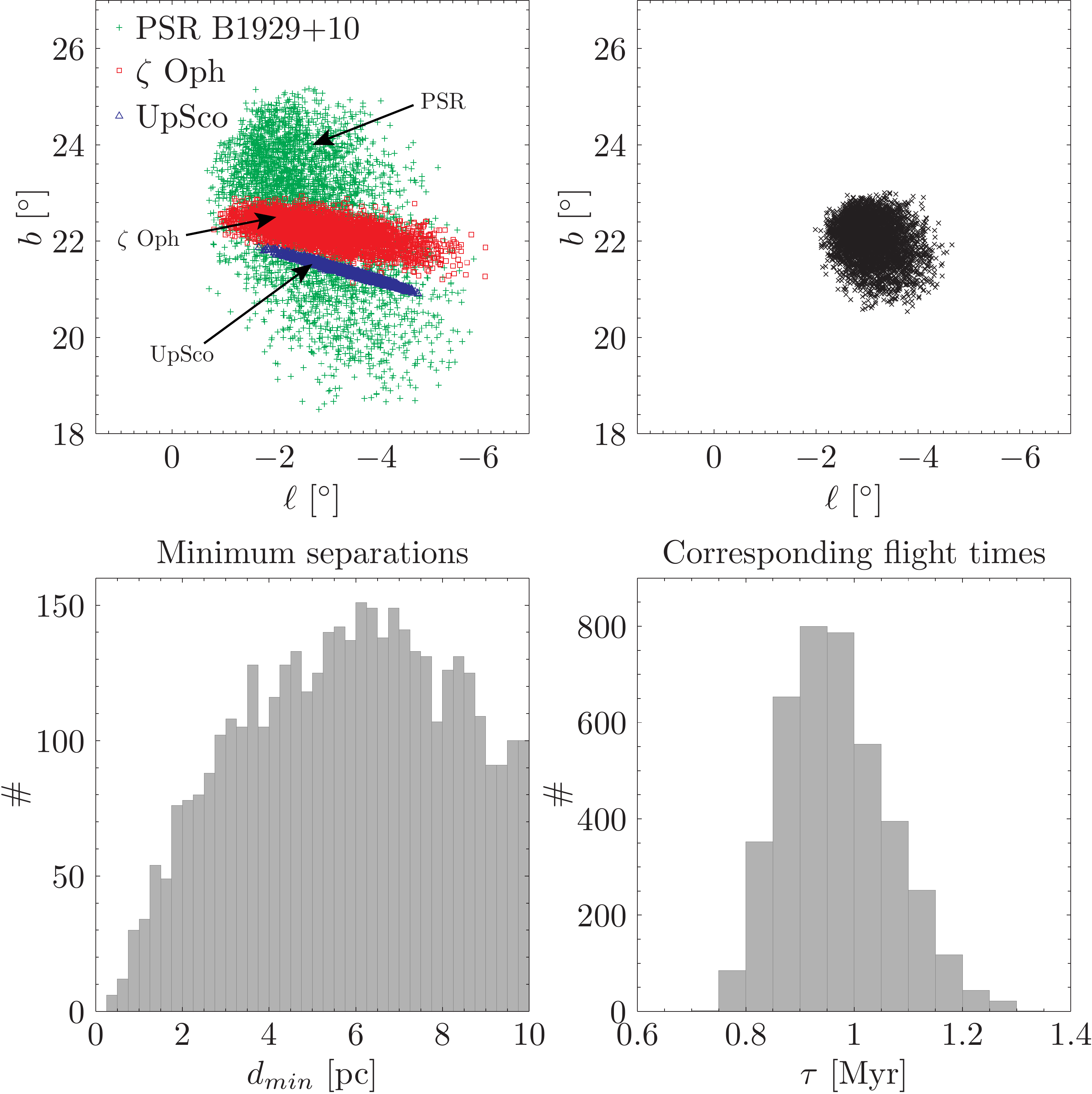}
\caption{Software test with original data of PSR\,B1929+10 and $\zeta$ Oph.
  Upper left: positions of 
PSR\,B1929+10 (green plusses), $\zeta$ Oph (red squares), and center of Upper Scorpius (UpSco, blue triangles) 
  at closest convergence within 10 pc (4066 cases in 3 million runs); 
  we confirm Ho00 and Ho01 
  with their old input values (case 1 in Table 2). 
  Upper right: for the 4066 cases of close approach, we show here the 4066 mean positions. 
  Lower left: minimum separations (left) between
  PSR\,B1929+10 and $\zeta$ Oph (4066 of 3 million runs, case 1 in Table 2):
  a close approach of less than 1 pc was possible. 
  Lower right: corresponding flight times peaking at $\sim 0.95$ Myr.  
  All calculated by us to test our software with the old input values from Ho00 and Ho01, 
  which were, however, revised since them.
With the newest astrometric data for pulsar and runaway, this scenario is ruled out
(cases 4-6 in Table 2).
}
\end{figure}

We would also like to mention that the measurement uncertainties of the new Gaia DR2 values
for parallax and proper motion of $\zeta$ Oph are $\sim 5$ times larger
compared to the new Hipparcos data reduction (van Leeuwen 2007),
which is due to the fact that its large brightness 
(V=2.56, G=2.46 mag) was a more severe problem for Gaia than for Hipparcos:
according to Lindegren (2018), Gaia DR2 data (Gaia 2018)
are reliable only for stars fainter than 3.6 mag.
Indeed, while the ground-based proper motion and parallax of $\zeta$ Oph
(as listed, e.g., in VizieR) are consistent with the newer Hipparcos results (van Leeuwen 2007),
they are inconsistent with Gaia DR2 data -- and the
ground-based proper motions of bright nearby stars 
should be credible -- even though having larger measurement uncertainties. 
Hence, for a star as bright as $\zeta$ Oph, the Gaia DR2 data are unreliable.
However, this problem affects only very few stars 
among the 56 best-known runaway stars (Ho00, Ho01),
which we studied here. 
Hence, we use the new Hipparcos data reduction (van Leeuwen 2007) for $\zeta$ Oph
and the other few stars brighter than 3.6 mag, but the Gaia DR2 data for all others.

All previously mentioned calculations were done with the Galactic potential 
similar or identical as in Ho00 and Ho01, see our Equ. 3.
Also, when using the Galactic potential according to Equ. 4, 
$\zeta$ Oph and PSR\,B1929+10 cannot have met for any of the parameter sets
with either Hipparcos or Gaia data and the new RV from Zehe et al. (2018).

We conclude that, in contrast to earlier claims (Ho00, Ho01), 
that there is no evidence that the runaway star $\zeta$ Oph and the pulsar PSR\,B1929+10 
were ejected from the same SN;
this was also concluded in previous studies by Chatterjee et al. (2004), Tetzlaff et al. (2010), 
and Kirsten et al. (2015) 
based on new proper motions and parallaxes,
but it gets more severe with the new RV from Zehe et al. (2018).
Hence, other solutions for their origins have to be found. 
This conclusion holds for both the Hipparcos astrometry (case 4) and
the newest Gaia DR2 astrometry for $\zeta$ Oph (cases 5 and 6 in Table 2),
even though the latter are not credible ($\zeta$ Oph too bright).
Given that the presumable SN releasing $\zeta$ Oph and PSR\,B1929+10 would have
been at $150 \pm 5$ pc (our calculation), its possible contribution to 
the $^{60}$Fe detected on Earth would have been marginal anyway.

\section{The new case: $\zeta$ Oph and PSR\,B1706-16}

To find systems of runaways and pulsars, we numerically integrated backward the flight paths 
of all ten well-known OB-type runaways from or near Sco-Cen-Lup (Ho00, Ho01) and 400 Galactic neutron stars 
with known distance 
and transverse velocity (Manchester et al. 2005) through the Galactic gravitational potential. 
We used Gaia DR2 data for most of the runaways (Gaia 2018), but Hipparcos data (van Leeuwen 2007) for those brighter 
than the Gaia limit (3.6 mag), including $\zeta$ Oph.
Since we concentrate here on Sco-Cen-Lup, we present input data and results on those ten runaway stars,
which come from somewhere near this association here, while results for other runaway stars will
be presented elsewhere. The following RV were used: 
\begin{itemize}
\item $\zeta$ Oph (O9.5): mean RV = $12.2 \pm 3.3$ km/s from 48 spectra in Zehe et al. (2018) --
this star is too bright for Gaia (V=G=2.56 mag), so that we use the Hipparcos data (van Leeuwen 2007).
\item $\kappa ^{1}$ Aps (B2): spectroscopic binary with center-of-mass RV = $16.5 \pm 5.0$ km/s from 7 spectra in Jilinski et al. (2010).
\item j Cen (B3): RV = $33.0 \pm 7.9$ km/s from spectra in Campbell (1928) and Evans (1979).
\item $\zeta$ Pup (O4I): $-25.0 \pm 5.0$ km/s from 6 spectra in Frost et al. (1926), Campbell (1928), Wilson (1953),
Gontcharov (2006), Barbier-Brossat et al. (1994) -- this star is too bright for Gaia (V=G=2.25 mag),
we use the Hipparcos data (van Leeuwen 2007).
\item HD 73105 (B3): mean RV = $22.0 \pm 3.0$ km/s from 6 spectra in Jilinski et al. (2010).
\item L Vel (B2): mean RV = $16.5 \pm 3.0$ km/s from 16 spectra in Jilinski et al. (2010).
\item HD 86612 (B5): mean RV = $19.4 \pm 0.8$ km/s from 5 spectra in Jilinski et al. (2010).
\item HD 88661 (B5): mean RV = $31.0 \pm 3.0$ km/s from 5 spectra in Thackeray et al. (1973).
\item V716 Cen (B5): spectrocopic binary with center-of-mass RV = $-10.3 \pm 6.9$ km/s from 27 spectra in Bakis et al. (2008).
\item HD 152478 (B3): mean RV = $19.7 \pm 9.5$ from Neubauer (1930), Buscombe \& Kennedy (1965), and Evans (1979).
\end{itemize}

After calculating backwards the above ten runaway stars and all 400 neutron stars to Sco-Cen-Lup
(but note that none of the other 46 runaway stars in Ho00 \& Ho01 come from Sco-Cen-Lup),
we found one convincing solution: the radio pulsar PSR\,B1706-16 and $\zeta$ Oph were at the same time 
at the same position: in 61921 of 3 million Monte Carlo runs, 
they came closer than 10 pc (in 649 runs within 1 pc). 
For runs with close approach within UCL, they happened at $107 \pm 4$ pc distance 
(Galactic coordinates $l=-13 \pm 3^{\circ}$ and $b=21 \pm 2^{\circ}$) $1.78 \pm 0.21$ Myr ago (flight time), 
for a pulsar RV of $260 \pm 43$ km/s. 
In the best case, for a pulsar RV of 265.5 km/s, an approach within less than 0.5 pc happened 1.58 Myr ago 
at 111 pc distance in UCL. 
See Figs. 2-4 and Tables 3-5.
At the location and time of the close approach, 
a SN in a binary star took place ejecting PSR\,B1706-16, $\zeta$ Oph, and $^{60}$Fe.
All close approaches of PSR\,B1706-16 and $\zeta$ Oph were in UCL,
while we can exclude UpSco and LCC as their place of origin.
For another certain pulsar RV range, the flight paths of PSR\,B1706-16 and $\zeta$ Oph
crossed in LCC, but at a time longer ago than the pulsar upper age limit, so that such a solution is excluded.

\begin{figure*}
\centering
\includegraphics[width=16.5cm]{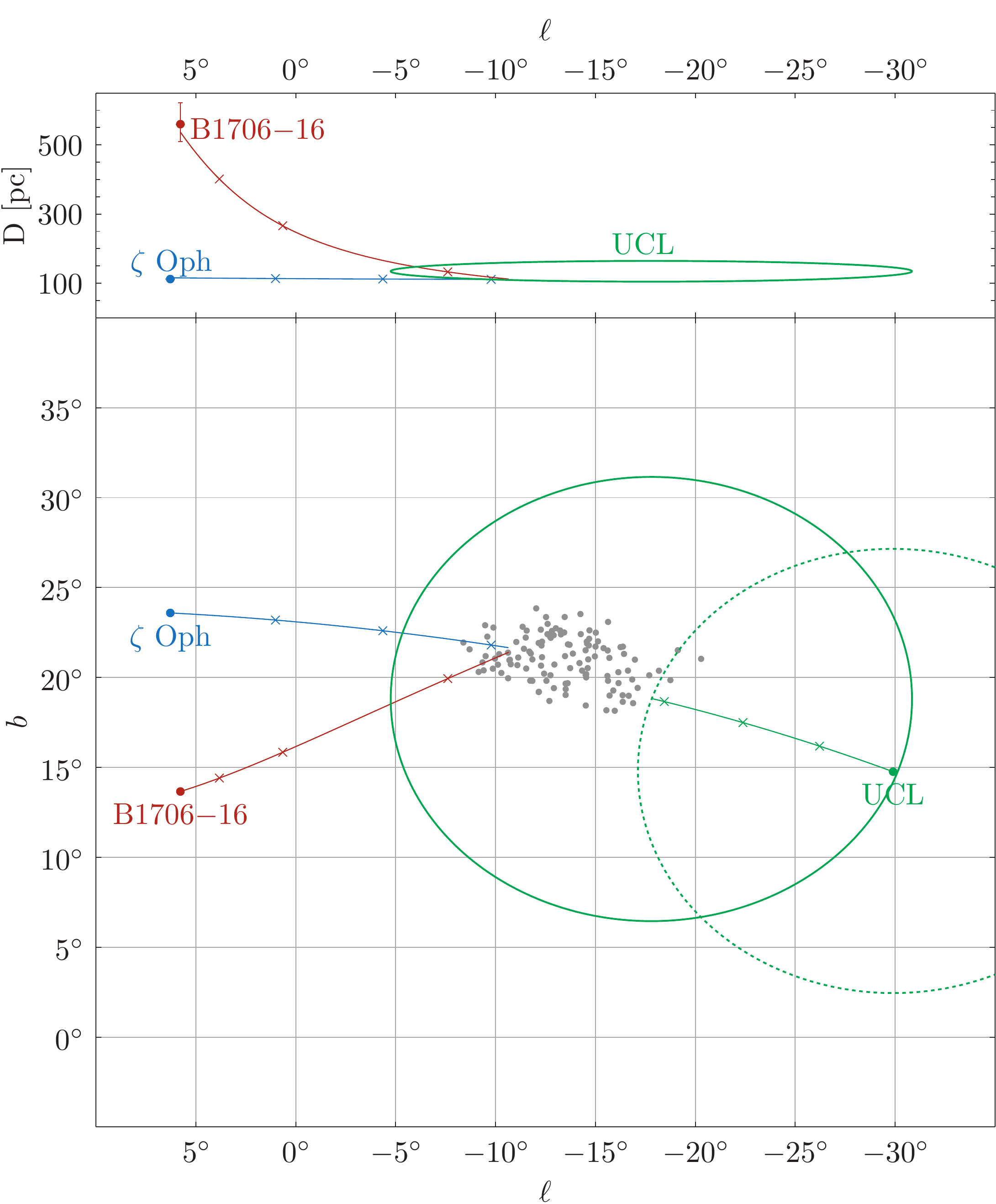}
\caption{Close approaches between PSR\,B1706-16 and $\zeta$ Oph.
Lower panel: For the closest approach between PSR\,B1706-16 and $\zeta$ Oph (best case)
projected on sky (Galactic longitude $l$ versus latitude $b$ at epoch of date), 
we show the flight paths
(coloured lines) starting at their current positions (left), marked by 
small crosses after each 0.5 Myr of flight time, ending 
1.58 Myr before present at the closest approach within 0.5 pc (center).
UCL is shown as full circle for 1.58 Myr ago and as dotted circle for the present time.
The grey dots show all 117 close approaches within 
10 pc inside UCL (Table 5), 
mean flight time $1.78 \pm 0.21$ Myr, mean distance at closest approach $107 \pm 4$ pc.  
Upper panel: distance $D$ versus Galactic longitude $l$ for the past flight 
paths (best case);
the lines from $\zeta$ Oph and the pulsar do not start 
exactly at their nominal positions (dots), because we show 
here the distances which were used as input in that 
particular run (best case among 3 million runs by varying the input 
using their measurement uncertainties). 
See also Tables 3-5 and Figs. 3 \& 4.}
\end{figure*}

Our scenario is consistent with all that is known about the runaway $\zeta$ Oph, 
the pulsar PSR\,B1706-16, and the UCL group: \\
The O9.2-9.5-type star $\zeta$ Oph has a current distance of $112.2 \pm 2.6$ pc (van Leeuwen 2007), 
and its 3D flight path intersected with UCL, which has the most massive stars 
among the three sub-groups: UCL without O-type stars, $\sim 20$ B1-2.5 stars, 
$\sim 50$ B3-9 stars, $\sim 70$ A stars (Pecaut \& Mamajek 2016), and many lower-mass stars. \\                                                                  
PSR\,B1706-16, at a current distance of $560 \pm 56$ pc, derived from its dispersion measure  
$24.891 \pm 0.001$ pc/cm$^{3}$ (Stovall et al. 2015), using the most recent model for the Galactic distribution 
of free electrons (Yao et al. 2017), has a characteristic spin-down age of 1.64 Myr (Hobbs et al. 2004), 
so that the flight time since the SN, $1.78 \pm 0.21$ Myr (or 1.58 Myr in the best case),
provides a credible solution. \\
UCL has a mean Gaia distance of $135.9 \pm 0.5$ pc (Wright \& Mamajek 2018) and an 
age of $\sim 10-20$ Myr.\footnote{10-20 Myr age range as found by different age determination techniques:
isochrones fitted to color-magnitude-diagram (Pecaut \& Mamajek 2016),
kinematic convergence point method (Wright \& Mamajek 2018), and
Lithium burning (Song et al. 2012); last consensus age $16 \pm 2$ Myr (Hyde \& Pecaut 2018);
the older upper age limit found by Fuchs et al. (2006) is based on the outdated Schaller et al. (1992)
theoretical isochrones; while Madsen et al. (2002) consider a 20-30 Myr age range, 
they finally conclude: {\it The combined sample again indicates some expansion, 
roughly consistent with a kinematic age of 20 Myr ... for the Scorpius-Centaurus complex}.}
Hence, $\zeta$ Oph could have been ejected from UCL by a SN in a binary $1.78 \pm 0.21$ Myr ago; 
indeed, the present mass function of all stars in UCL compared to the IMF predicts about one O9-type 
star: $\zeta$ Oph. For the neutron star PSR\,B1706-16, this scenario works for RV$=260 \pm 43$ km/s, 
typical for pulsars (Hobbs et al. 2005, Verbunt et al. 2017). 

While we used the proper motion of PSR\,B1706-16 from Fomalont et al. (1997)
as obtained from four positons with 10 year epoch difference with the high-precision VLA array,
Zou et al. (2005) and Jankowski et al. (2019) determined its proper motion
by the pulsar timing technique.
All proper motion in right ascension are fully consistent within $1 \sigma$;
the proper motion in declination in Zou et al. (2005) and Fomalont et al. (1997)
are consistent within $1 \sigma$, while the value in Jankowski et al. (2019) is deviant by $2 \sigma$. 
However, proper motions from radio timing are unreliable in case of large timing noise --
and indeed, PSR\,B1706-16 has a non-zero 2nd derivative of the frequency,
$0.9 \pm 0.2 \cdot 10^{-24} s^{-3}$, 
e.g. due to glitches (Jankowski et al. 2019). 
The observations in Zou et al. (2005) and Jankowski et al. (2019) have only
1-3 yr epoch difference, and Zou et al. (2005) show (their figure 2) that
in particular the right ascension and, hence, proper motion in declination
have strong uncertainties for such short epoch differences.
We therefore consider the Fomalont et al. (1997) proper motion the most reliable.
Furthermore, previous dispersion measure distances for PSR\,B1706-16 are fully consistent
with the current best value, but have larger error bars (e.g. Jankowski et al. 2019).

To estimate the significance of a certain number of close meetings between $\zeta$ Oph and the pulsar PSR\,B1706-16 within UCL
in the past (e.g. 117 within 10 pc as we found above),
we simulated virtual pulsar-runaway-pairs with parameters and uncertainties as for
$\zeta$ Oph and PSR\,B1706-16:
position and 3D kinematics of $\zeta$ Oph trace it back to UCL, for a certain
time range (1 to a few Myr), so that a close approach of $\zeta$ Oph with a pulsar 1 to few Myr ago
will always be in UCL; we have to calculate here only the number of runs producing a close approach.
For virtual SNe in binaries inside UCL, we created virtual pulsar-runaway-pairs.
We ran them forward with the kinematic properties (proper motions and RVs) as obtained 
before by tracing back (Tables 3 and 4, Fig. 4), including the RV for the pulsar of $260 \pm 43$ km/s.
We did such calculations for flight times from 0.5 to 3.0 Myr in steps of 0.25 Myr --
this gives us 11 points in time, where runaway and pulsar could be located at that time
after having been ejected from the virtual SN.
For each of these 11 cases, we then traced back the runaway and the pulsar starting
from their virtual positions and now using the kinematic properties (proper motions and RVs)
as obtained before by tracing back (Tables 3 and 4) -- and varrying them within their
measurement uncertainties (as well as the parallax, as usual) for 3 million trials each 
(and by drawing the RV of the pulsar from the Hobbs et al. distribution as usual).
For each such trial, we can then obtain as usual the minimum distance between runaway and pulsar etc.
This procedure thus yields the number of expected close approaches (within e.g. 1 or 10 or 15 pc).
We display these values in Fig. 5.
As a result,
we obtained (and, thus, expect) close meetings within 10 pc inside UCL in six of 3 million runs.
Within $95\%$ confidence interval under the assumption of binomial distribution, we would expect 3 to 14 close approaches;
or, e.g., 58 close approaches within 15 pc ($95\%$ confidence: 45-75), 235 within 20 pc ($95\%$ confidence: 207-267), etc. --
and since we found 117 close approaches in 3 million runs within 10 pc inside UCL by back-tracing, 
our solution is compatible with the scenario that the two objects were ejected from the same position.

Also, we simulated random pulsars and random runaway stars with parameters
sampling those of the 400 neutron stars (ATNF etc.) and 56 runaway stars (Ho00, Ho01);
for the random neutron stars, we have drawn the RV from the empirical distribution
of space velocity (Hobbs et al. 2005) as for the real neutron stars.
We then performed many Monte Carlo simulations of such unrelated pulsar-runaway-pairs:
we would need $4 \cdot 10^{10}$ trials to get 120 cases, where such random pulsar-runaway-pairs can 
meet in a time range 1.5-2.0 Myr ago.
Hence, our 117 cases in 3 million trials for the real pulsar-runaway-pair cannot be a chance coincidence,
the probability for such a case would be only $7.5 \cdot 10^{-5}$ (the fraction of 3 million to $4 \cdot 10^{10}$ runs). 

Based on previous, less precise data, it was claimed
that $\zeta$ Oph and PSR\,B1929+10 were ejected by a SN (Ho00, Ho01): 
there were 30,822 of 3 million runs with a close approach
within 10 pc, but this large number
of encounters was obtained with restricted input parameters for pulsar parallax and RV.
We did calculations for our case ($\zeta$ Oph and PSR\,B1706-16) with similar 
restrictions and then got 146,727 of 3 million cases with close encounters $\le 10$ pc;
hence, our solution is more probable
than the one by Ho00 \& Ho01.

The location of our closest approach is inside a cavity in the interstellar medium (Table 3):
the current location of a cavity in the interstellar medium (ISM) found by Capitanio et al. (2017)
is considered to be consistent with the predicted location of SN 16 (from B16),
as discussed in Capitanio et al. (2017);
this would then indicate a $\sim 20\%$ measurement uncertainty for both the cavity and the SN predictions.
Note that the SN would not need to have happened in the very center of the present cavity,
because the interstellar medium density within it probably was inhomogeneous.

There is also an Integral SPI 1808 keV $^{26}$Al $\gamma$-ray source (0.7 Myr half-life, see Table 3) 
in Sco-Cen-Lup, but it is probably related to a SN in UpSco, not to our SN in UCL.
The SN in UpSco related to the $\gamma$-ray source may have ejected the neutron star 
RXJ\,1856.5-3754 (Walter et al. 1996)
probably originated in a SN $\sim 0.5$ Myr ago (Tetzlaff et al. 2010, 2011b),
also seen as a large cavity in HI (Krause et al. 2018);
the radio-quiet X-ray pulsar RXJ\,1856.5-3754 is now at $\sim 123$ pc distance,
detected in the optical (Walter et al. 2010).  

\begin{figure}
\centering
\includegraphics[width=\columnwidth]{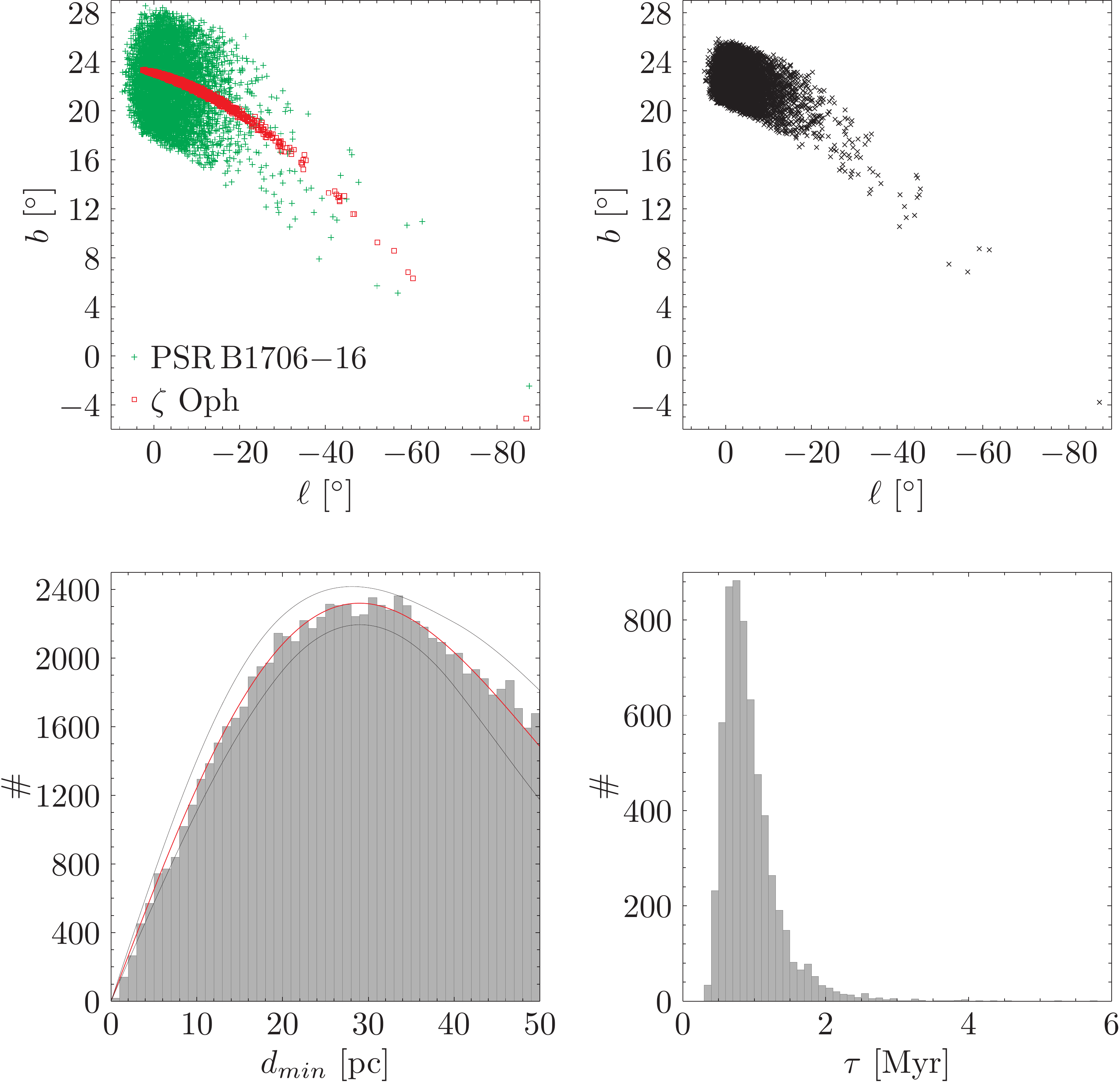}
\caption{Close approaches between PSR\,B1706-16 and $\zeta$ Oph:
upper left: 
positions in Galactic longitude l and latitude b at closest convergence 
of PSR\,B1706-16 (green plusses) and $\zeta$ Oph (red squares) within 10 pc (61,921 of 3 million runs). 
Upper right: mean position for each of those 61,921 pairs (PSR\,B1706-16 and $\zeta$ Oph). 
Bottom left: we fitted this distribution of minimum separations 
(up to 50 pc, 300,000 runs only) to equation A3 in Ho01
and obtained as mean minimum separation $\mu = 14.3$ pc with standard deviation $\sigma = 13.8$ pc
(red curve with enveloping grey curves) -- consistent with a very close approach ($\simeq 0$ pc).
Given the large number of runs (3 million) with varying input parameters,
the number of runs with a small minimum separation is expected to be small (see text).
Bottom right: corresponding flight times for those 61,921 cases with $\le 10$ pc separation. 
}
\end{figure}

\begin{figure*}
\centering
\includegraphics[width=18cm]{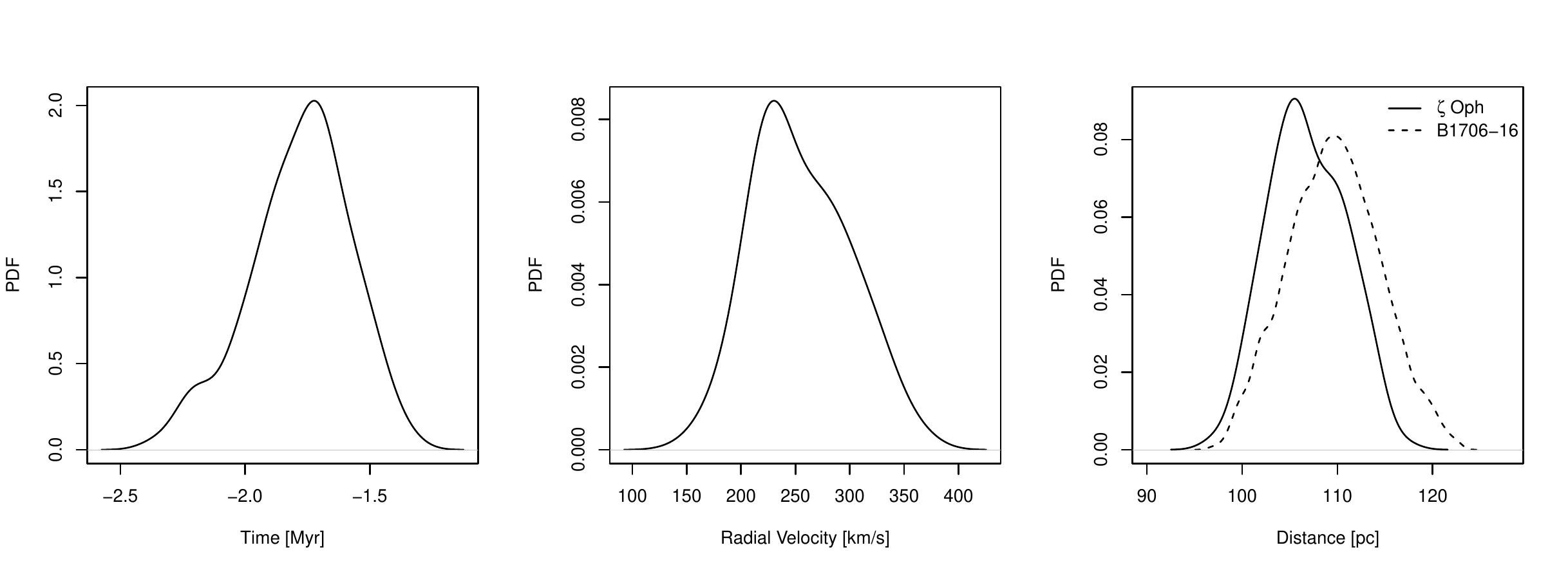}
\caption{Distributions of flight time, pulsar RV, and SN distance:
For the 117 best cases of close approaches within 10 pc between
$\zeta$ Oph and PSR\,B1706-16 inside UCL, we present here the probability
density functions (PDF) for the flight time since the supernova (left), 
the radial velocity of the pulsar used (middle), and the
distances of $\zeta$ Oph and PSR\,B1706-16 at the time of the closest approaches (right).
The mean values are listed in Table 5.}
\end{figure*}

\begin{table*}
\caption{Input values for calculating backward the motion of $\zeta$ Oph and PSR\,B1706-16}
\begin{tabular}{lcccccc} \hline
Object & \multicolumn{2}{c}{Galactic Coordinates} & Distance & \multicolumn{2}{c}{Proper motion or U,V [km/s]} & Rad. Vel. \\
       & longitude l & latitude b  & d    & $\mu _{\alpha} \cdot \cos \delta$ & $\mu _{\delta}$  & or W \\ 
       & [$^{\circ}$] & [$^{\circ}$] & [pc] &                 & & \\ \hline
\multicolumn{7}{l}{Nominal values for runaway star, pulsar, and associations (present-day coordinates):} \\ \hline
$\zeta$ Oph (a)        & 6.28123      & 23.58774  & $112.2^{+2.6}_{-2.5}$  & $15.26 \pm 0.26$ mas/yr & $24.79 \pm 0.22$ mas/yr & $12.2 \pm 3.3$ km/s \\
PSR\,B1706-16 (b)      & 5.77         & 13.66     & $560 \pm 56$     & $3 \pm 9$ mas/yr & $0 \pm 14$ mas/yr & (c) \\ 
UCL center (d)       & $-29.1$      & 14.8        & $135.9^{+0.5}_{-0.4}$  & U=-$5.9 \pm 2.0$ km/s & V=-$20.0 \pm 0.7$ km/s & W=-$5.8 \pm 1.3$ km/s \\ 
LCC center (d)       & $-59.6$      & 8.3         & $115.2 \pm 0.3$  & U=-$9.0 \pm 1.9$ km/s & V=-$20.6 \pm 0.9$ km/s & W=-$6.3 \pm 0.5$ km/s \\ 
UpSco center (d)     & $-8.9$       & 19.4        & $143^{+0.3}_{-0.4}$  & U=-$6.2 \pm 1.6$ km/s & V=-$16.9 \pm 1.1$ km/s & W=-$7.1 \pm 2.2$ km/s \\ \hline \hline
\multicolumn{7}{l}{Other relevant objects to compare with (present-day coordinates):} \\ \hline
$^{26}$Al source (e)   & $-10 \pm 10$ & $20 \pm 10$ & \multicolumn{3}{l}{6$\sigma$ detection at UpSco (e)} \\
ISM cavity (f)  & $-18 \pm 6$  & $18 \pm 6$  & $125 \pm 25$ & (g) & (g) & (g)  \\ 
pred. SN 16 (h)    & $-17 \pm 6$  & $25 \pm 6$  & $96 \pm 19$  & (g) & (g) & (g)  \\ \hline \hline
\multicolumn{7}{l}{Input parameters for the best case (close approach within 0.5 pc, present-day coordinates):} \\ \hline
$\zeta$ Oph                & 6.28123     & 23.58774    & 115.6 & 15.38 mas/yr & 24.85 mas/yr & 4.6 km/s (i) \\
PSR\,B1706-16              & 5.77        & 13.66       & 537   & 6.87 mas/yr & 2.72 mas/yr & 265.5 km/s \\
UCL center (d)             & $-29.1$     & 14.8        & 137   & U=$-4.87$ km/s & V=$-19.94$ km/s & W=$-5.56$ km/s \\ \hline  \hline
\multicolumn{7}{l}{Mean input parameters for 117 best cases (close approaches within 10 pc in UCL, present-day coordinates, note j):} \\ \hline
$\zeta$ Oph                & 6.28123   &  23.58774  & $115 \pm 10$ & $15.26 \pm 0.27$ mas/yr & $24.81 \pm 0.24$ mas/yr & $6.7 \pm 2.2$ km/s \\
PSR\,B1706-16              & 5.77      &  13.66     & $564^{+70}_{-56}$ & $6.01 \pm 1.2$ mas/yr & $2.9 \pm 1.1$ mas/yr & $260 \pm 43$ km/s \\
UCL center (d)             & $-29.1$   &  14.8      & $135.85 \pm 0.5$ & U=-$4.1 \pm 2.0$ km/s & V=-$20.09 \pm 0.69$ km/s & W=-$5.6 \pm 1.3$ km/s \\ \hline
\end{tabular}

Notes: 
(a) Hipparcos (van Leeuwen 2007), new RV (Zehe et al. 2018);
Gaia DR2 unreliable for $\zeta$ Oph (Lindegren 2018). 
(b) Current coordinates from ATNF (Hobbs et al. 2004, Manchester et al. 2005),
parallax from the dispersion measure $24.891 \pm 0.001$ pc cm$^{-3}$ (Stovall et al. 2015),
converted with the most recent model (Yao et al. 2017)
for the Galactic distribution of free electrons, 
proper motion from Fomalont et al. (1997);
pulsar discovered by Large et al. (1969). 
(c) RV unknown, Monte Carlo simulation with 3 million runs drawing velocity from 
Maxwellian distribution (Tetzlaff et al. 2010).  
(d) from Wright \& Mamajek (2018) with a radius of $\sim 30$ pc for UCL ($\pm 13^{\circ}$), $\sim 23.25$ pc for LCC,
and 17 pc for UpSco, 1.5 times the scale length of the half-mass radius (Wright \& Mamajek 2018). 
(e) Diehl et al. (2010), 
(f) Current location of a cavity in the interstellar medium (ISM) found by Capitanio et al. (2017).
(g) These positions participate in the Galactic rotation, but have no other peculiar velocities.
(h) Predicted present-day parameters (B16) for SN 16 with $\sim 20\%$ uncertainties.  
(i) While this input RV of $\zeta$ Oph is $2 \sigma$ off the mean value (Zehe et al. 2018), 
the input, output, and total log likelihoods of this case are still the highest,
and the approach between runaway and pulsar is the smallest -- hence, our {\it best case}. 
A RV value of $\sim 4.6$ km/s was actually published by Garmany et al. (1980).
The inclination of the bow shock around $\zeta$ Oph also indicates motion
almost in the plane of the sky, i.e. with a very small RV (Gvaramadze et al. 2012, dell Valle \& Romero 2012),
even though an exact value for RV cannot be determined from the bow shock alone,
because it is quite irregular shaped. 
(j) For the 117 cases of close approaches within 10 pc inside UCL
(after tracing back $\zeta$ Oph, PSR\,B1706-16, and UCL),
always unimodal distributions (Fig. 4),
we present here the mean input values with $1 \sigma$ error bars (l,b fixed). 
\end{table*}

The SN position 
from tracing back $\zeta$ Oph and PSR\,B1706-16 is at 
Galactic longitude $-16 \pm 4^{\circ}$ (i.e. $344 \pm 4^{\circ}$)
and latitude $15 \pm 3^{\circ}$ (present-day epoch, Table 4),
which is in an area of UCL 
with $\sim 12-15$ Myr age (Pecaut \& Mamajek 2016, their figure 9). 
The difference between this age and the flight time is the lifetime of PSR\,B1706-16's progenitor, 
which therefore had a main-sequence mass of $\sim 15-19$ M$_{\odot}$ (Ekstr\"om et al. 2012), 
indeed more massive than $\zeta$ Oph with $\sim 13 \pm$ few M$_{\odot}$ (Marcolino et al. 2009)
and $\sim 16 \pm$ few Myr lifetime (Ekstr\"om et al. 2012). 
If Black Holes are the end-states of stars originally more massive than $\sim 18$ M$_{\odot}$ (Smartt 2015), 
then we can restrict the SN progenitor mass to $\le 18$ M$_{\odot}$.

\begin{table}
\caption{Predicted Supernova position from tracing back $\zeta$ Oph and PSR\,B1706-16}
\begin{tabular}{lcc} \hline
\multicolumn{3}{l}{for best case approach within 0.5 pc inside UCL:} \\ 
       & \hspace{-1cm} {\small 1.58 Myr ago} & {\small present day} \\ \hline
Galactic longitude l & $-10.7^{\circ}$ & $-18.9^{\circ}$  \\
Galactic latitude b  & $21.55^{\circ}$ & $13.4^{\circ}$   \\
Distance from Earth  & 111.2 pc        & 107.9 pc         \\ 
\multicolumn{3}{l}{(for pulsar radial velocity 265.5 km/s)} \\ \hline
\multicolumn{3}{l}{for 117 best case approaches within 10 pc inside UCL:} \\ 
       & \hspace{-1cm} {\small $1.78 \pm 0.21$ Myr ago} & {\small present day} \\ \hline
Galactic longitude l & $-13 \pm 3^{\circ}$ & $-16 \pm 4^{\circ}$  \\
Galactic latitude b  & $21 \pm 2^{\circ}$  & $15 \pm 3^{\circ}$   \\
Distance from Earth  & $107 \pm 4$ pc      & $109 \pm 5$ pc       \\ 
\multicolumn{3}{l}{(for pulsar radial velocity $260 \pm 43$ km/s)} \\ \hline
\end{tabular}

Note: More details and input parameters in Tables 3 \& 5.
\end{table}

The pre-SN system became unbound if at least half of its total pre-SN mass was ejected. 
For a neutron star with the usual $\sim 1.4$ M$_{\odot}$ and $\zeta$ Oph 
with $\sim 13$ M$_{\odot}$ (Marcolino et al. 2009), the final SN progenitor had $\ge 16$ M$_{\odot}$,
and $\le 18-19$ M$_{\odot}$ as found above. 
Given the peculiar space velocity of $\zeta$ Oph (19.7 km/s) as velocity on a circular orbit 
around the center-of-mass of the progenitor binary, and given the mass ratio between runaway 
and SN progenitor, the latter had an orbital velocity of 14-16 km/s; 
the relative velocity then yields a pre-SN binary separation of 23-25 au and $\zeta$ Oph's 
orbital period of $\sim 20-22$ yr. High rotational velocity (near breakup)
of $\zeta$ Oph (Marcolino et al. 2009) may indicate mass exchange, so that the orbit should have been highly eccentric --
the Roche lobe radius is $\sim 1$ au; detection of SN debris in $\zeta$ Oph's atmosphere would be 
challenging due to strong rotational line broadening. 
Since the runaway star $\zeta$ Oph and the pulsar PSR\,B1706-16 now fly to similar directions (Fig. 2), 
the SN kick on the pulsar was opposite its orbital direction, 
but did unbind the system: 
for a kick of $253 \pm 54$ km/s in the opposite direction 
from present space velocity plus former orbital velocity with $70 \pm 6^{\circ}$ 
3D angle between the current runaway and pulsar paths, 
the runaway velocity should be moderate (Renzo et al. 2019), as is the case for $\zeta$ Oph.

\begin{table*}
\caption{Output values for close approaches $\zeta$ Oph and PSR\,B1706-16}
\begin{tabular}{lllll} \hline
Object             & longitude l [$^{\circ}$] & latitude b [$^{\circ}$] & distance [pc] & notes \\ \hline
ISM cavity center  & $-16 \pm 6$    & $20 \pm 6$ & $144 \pm 26$  & (a) for 1.58 Myr ago \\ \hline
predicted SN 16    & $-11 \pm 6$    & $27 \pm 6$ & $115 \pm 19$  & (b) for 1.58 Myr ago \\ \hline \hline
\multicolumn{5}{l}{best case approach within 0.5 pc, for epoch 1.58 Myr ago:} \\ \hline
$\zeta$ Oph        & $-10.681$      & 21.653     & 111.1      & Fig. 2  \\
PSR\,B1706-16      & $-10.802$      & 21.466     & 111.2      & Fig. 2  \\
UCL center (c)     & $-17.8$        & 18.8       & 134.5      & RV pulsar $265.5$ km/s   \\  \hline \hline
\multicolumn{5}{l}{means for 117 best cases of close approaches within 10 pc within UCL, for respective epoch (d):} \\ \hline
$\zeta$ Oph        & $-13.4 \pm 2.3$ & $21.1 \pm 0.5$ & $106.6 \pm 3.8$ & flight time $1.78 \pm 0.21$ Myr \\
PSR\,B1706-16      & $-13.5 \pm 3.1$ & $21.0 \pm 2.5$ & $107.6 \pm 4.0$ & RV pulsar $260 \pm 43$ km/s \\ \hline
\end{tabular}

Notes: 
(a) From Capitanio et al. (2017), 
position here for 1.58 Myr ago just considering 
differential Galactic rotation for Galactic potential of Model III in in Bajkova \& Bobylev (2017). 
(b) B16, see our Sect. 5. 
(c) With a radius of $\sim 30$ pc for UCL ($\pm 13^{\circ}$), 
1.5 times the scale length of the half-mass radius (Wright \& Mamajek 2018). 
(d) For the 117 cases of close approaches within 10 pc inside UCL
(after tracing back $\zeta$ Oph, PSR\,B1706-16, and UCL),
we present here the mean output values with $1 \sigma$ error bars (Fig. 4).  
\end{table*}
 
None of the other 55 runaways (Ho00, Ho01) can be traced back to a close approach with a 
pulsar within Scorpius-Centaurus-Lupus, but the flight paths of some of them crossed it:
among the runaway stars, in addition to $\zeta$ Oph,
we noticed that the flight paths of $\kappa ^{1}$ Aps (HIP 76013) and j Cen (HIP 57669) crossed 
LCC, UCL, and UpSco some 2.9-0.5 Myr ago, first in LCC, then UCL, then UpSco,
where they may have been ejected. For j Cen, it was suggested to originate from the cluster IC 2602 
(Ho01) based on Hipparcos data, but that is excluded from Gaia DR2 data;
for $\kappa ^{1}$ Aps, Ho01 suggested that it was ejected from LCC 2-3 Myr ago,
based on Hipparcos data.
With Gaia DR 2 data, at the closest approach of their past flight paths to Earth, $\kappa ^{1}$ Aps and j Cen 
were at distances of 157 pc and 127 pc, respectively (1.8 and 2.9 Myr ago), $\kappa ^{1}$ Aps in UCL, j Cen in LCC --
but without any neutron stars found at their respective positions in the past.
Both $\kappa ^{1}$ Aps and j Cen have high rotational velocities,
$v \cdot \sin i = 250-300$ km/s and 230-270 km/s, respectivcely (Bernacca \& Perinotto 1970, Uesugi \& Fukuda 1970,
Glebocki \& Gnacinski 2005), possibly indicating previous mass transfer.
Since their flight time from UCL/LCC is much smaller than the association age (10-20 Myr), a SN ejection
may be more likely than a dynamical ejection, which happens very early in the evolution.

The seven other runaway stars listed above just came close to Scorpius-Centaurus-Lupus.
According to Ho01, also HD 73105 (HIP 42038) came from UCL (or IC 2391),
HD 86612 (HIP 48943) from LCC, and V716 Cen (HIP 69491) from UCL (or Ceph OB6),
which we cannot confirm:
for all three of them, Ho01 used different RV data (and Hipparcos instead of Gaia DR2),
as already pointed out by Jilinski et al. (2010),
e.g. the latter (V716 Cen) was found to be a spectroscopic binary.

Furthermore, given their parallaxes and proper motions, in addition to PSR\,B1706-16
and RXJ\,1856-3754, the following neutron star may have originated 
from Scorpius-Centaurus-Lupus (for some RV between $-1000$ and $+1000$ km/s):
B0203-40, B0329+54, B0626+24, B1114-41, B1159-58, J1548-5607, B1804-08, and B1919+21
(but no matching runaway stars were found).
In addition, a particular intersting case may be J1548-5607, a 0.25 Myr young pulsar
(i.e. too young to have contributed to $^{60}$Fe on Earth),
now located at $2.6 ^{+2.8} _{-1.7}$ kpc distance (our Table 1),
but projected on sky just at the southern egde of UCL and moving south;
for a large RV of $1200-1400$ km/s, it could have been $30-40$ pc off
the UCL center $\sim 0.25$ Myr ago (but no runaway star found).

\begin{figure}
\centering
\includegraphics[width=8cm]{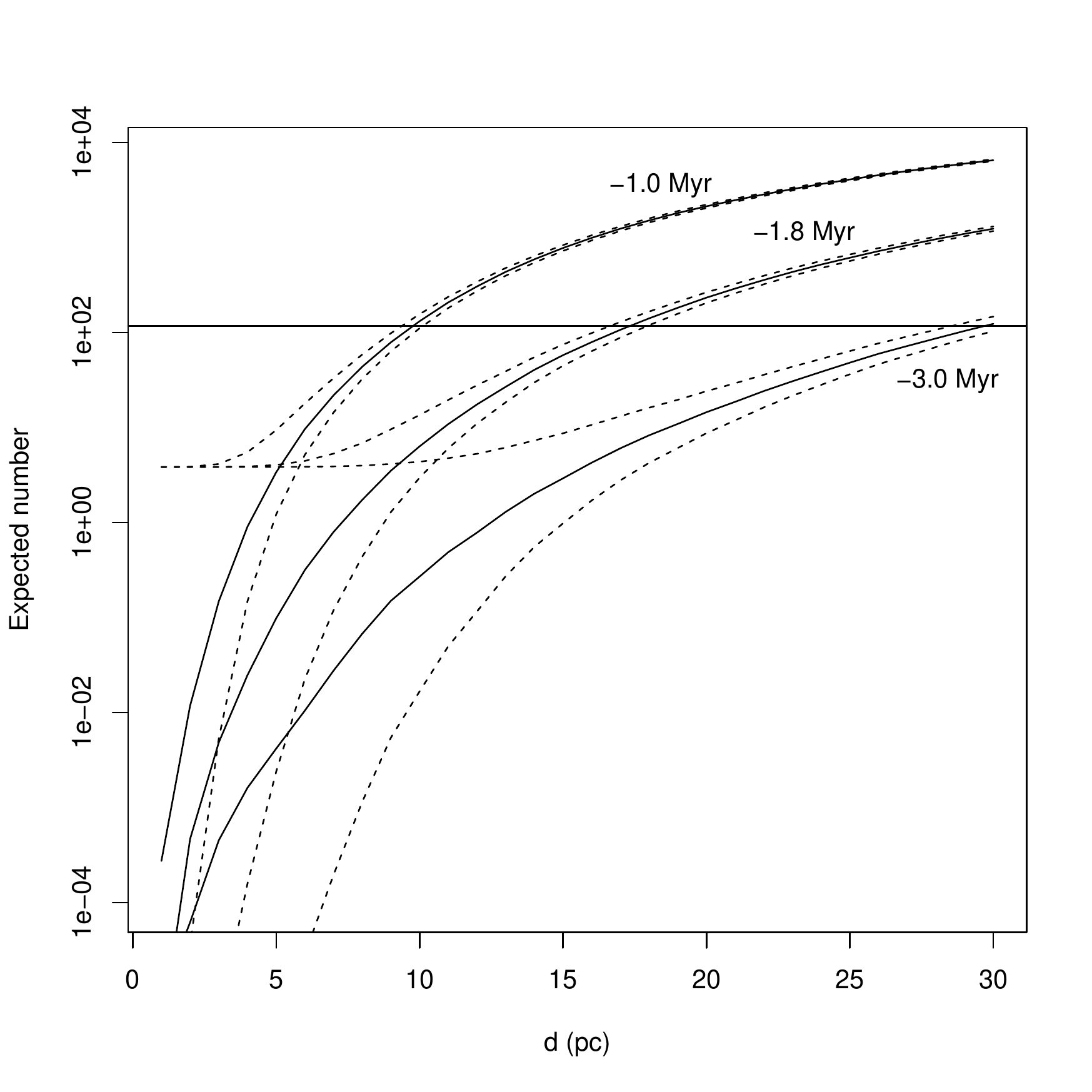}
\caption{Expected number of close approaches
versus minimum separation $d$ (in pc) achieved in our simulations
(see text Sect. 4) for three different flight times: 1.0 (top), 1.8 (middle), and 3.0 Myr (bottom).
The solution presented by us with a flight time of $1.78 \pm 0.21$ Myr (the time of the SN),
where we found 117 close approaches within 10 pc (within UCL),
is indicated by the green horizontal line.
For a simulated flight time of 1.8 Myr, we expect six close approaches within 10 pc in 3 million trails
plotted as black curve -- with $95\%$ confidence interval from 3 to 14 cases indicated by dashed black curves.
The number of close approaches found (117) is much larger than 3-14, so that our result is highly credible.
}
\end{figure}

\section{Discussion of the Breitschwerdt et al. (2016) model}

We determined above, that the final SN progenitor had a mass of $\ge 16$ M$_{\odot}$ 
and $\le 18-19$ M$_{\odot}$, and that it exploded $1.78 \pm 0.21$ Myr ago.
We notice that distance, position, and timing of this SN (our 117 best cases) would be consistent
with SN 16 in the B16 model (Table 5), but not the progenitor mass --
B16 gave 8.81 M$_{\odot}$ as progenitor mass for their SN 16,
which should have happened 1.5 Myr ago (B16).

However, there are several problems in the B16 model,
apart from the fact that B16 assumed that all 16 missing stars exploded as SN, 
while one or two most massive ones could have formed Black Holes, and a few should be ejected runaways: \\ 
(i) The most likely masses of the missing stars fix the lifetimes until their SNe.
However, because the association age of $\sim 22.5$ Myr assumed in B16 
is higher than $\sim 10-20$ Myr (see footnote 1),
the lower-mass SN progenitors, up to roughly 13-14 M$_{\odot}$ (Ekstr\"om et al. 2012), have not exploded, yet, 
even for a large age spread: there is no evidence that the lowest-mass SN progenitors 
would form first (Pecaut \& Mamajek 2016, Hyde \& Pecaut 2018). 
The more massive stars exploded in the last $\sim 6-7$ Myr yielding sufficient $^{60}$Fe, as detected on Earth.
Any uncertainty in the association age of at least a few Myr will propagate to all SN times predicted. \\
(ii) Even if 22.5 Myr would be the true UCL age,
stars below $\sim 11$ M$_{\odot}$ would not have exploded, yet: 
Fuchs et al. (2006) and B16 (as well as Schulreich et al. 2017 for a similar study of the Loop I bubble)
use a fit to the Schaller et al. (1992) isochrones: 
$\tau = 160$ Myr $\cdot M^{-0.932}$ with mass M in M$_{\odot}$ (Fuchs et al. 2006).
However, this fit
is inconsistent with the Schaller et al. (1992) isochrones at the lowest SN progenitor masses:
e.g. at 9 M$_{\odot}$, one would obtain 21 Myr from the above fit (as used in Fuchs et al. 2006, B16, and Schulreich et al. 2017), 
but 29-32 Myr from the original publication in Schaller et al. (1992);
Wallner et al. (2016) give $\sim 27$ Myr as life-time for such a progenitor mass in their table S8.
The above fit is valid only above $\sim 12$ M$_{\odot}$ (Schaller et al. 1992).
For new lifetime models with stellar rotation (Ekstr\"om et al. 2012),
the mass of stars with 22.5 Myr lifetime is $\sim$11 M$_{\odot}$. \\
(iii) There are further uncertainties in the B16 model like the choice of the 
mass-luminosity relation to obtain stellar masses and then the IMF itself:
Fuchs et al. (2006), B16, and Schulreich et al. (2017) choose the IMF from Massey et al. (1995) 
with an exponent $\Gamma = -1.1$, 
but there are also those by Salpeter (1955) with $-1.35$,
Miller \& Scalo (1979) and Kroupa (2001) with $-1.3$ (Chabrier 2003: $-1.3 \pm 0.3$).
With the most commonly used Salpeter IMF instead of the more rarely used Massey IMF,
we would expect a few less missing stars. \\
(iv) If $59 \pm 15\%$ of massive stars are multiple (Sana et al. 2012, Chini et al. 2012),
and if the masses within multiples are not very different 
(Pinsonneault \& Stanek 2006, Moe \& Di Stefano 2017, Zapartas et al. 2017b, Renzo et al. 2019),
and given that $67 \pm 17\%$ of massive multiple stars produce a runaway star in a core-collapse SN 
(Renzo et al. 2019, similar numbers in Zapartas et al. 2017b and Kochanek et al. 2019),
we would expect that at least one third of the missing stars are runaway stars,
maybe j Cen and/or $\kappa ^{1}$ Aps, and certainly $\zeta$ Oph --
in case of a flat mass ratio distribution, then somewhat less.
Some massive stars could also be missing due to dynamical ejection (Poveda et al. 1967, Ho00, Ho01).
Note that some ejected massive former companions of SNe move  with slow space velocities (those 
with wide separations during the SN, 'walkaways', Renzo et al. 2019), they may still be inside UCL/LCC. 
Including additional ejected runaways like $\zeta$ Oph
in the present UCL/LCC mass function will affect the predicted number, masses,
and lifetimes of missing stars, Hyde \& Pecaut (2018) assumed 3 massive runaways from UCL and LCC each. \\
(v) While according to B16 all UCL stars with $\ge 8.8$~M$_{\odot}$
would already have exploded, Hyde \& Pecaut (2018) lists 9 UCL members with 9-13~M$_{\odot}$,
of which in the list of UCL members in Fuchs et al. (2006), probably used in B16, 
only the high-proper-motion B2-star $\beta$ Lup 
and the B1-star $\mu^{1}$ Sco are missing -- B16 assigned lower masses than, e.g., Hyde \& Pecaut (2018).
That no significant amount of $^{60}$Fe arrived on Earth for the last $\sim 1.5$ Myr does not mean 
that SN activity in UCL/LCC ceased, 
as the time between two SNe is very roughly $\sim 1$ Myr -- $^{60}$Fe 
from a more recent SN might be on its way to Earth or another SN will happen soon;
e.g. RXJ\,1856.5-3754 might have originated in UpSco $\sim 0.5$ Myr ago, but too distant ($\sim 150$ pc,
Tetzlaff et al. 2010) and maybe also too recent for $^{60}$Fe to be detectable on Earth 
given the travel time and distance. 
The Lupus Loop (G330.0+15.0, $\sim 150-500$ pc) in the center of UCL may be the SN remnant from the latest SN in UCL
$\sim 15-31$ kyr ago (Ferrand \& Safi-Harb 2012), 
also too distant to bring detectable $^{60}$Fe to 
Earth and too recent to have reached Earth by now. \\
(vi) B16 and also Schulreich et al. (2017) neglected 
mass transfer in multiple stars,
which is, however, very important for their evolution
(Sana et al. 2012, Zapartas et al. 2017a). \\
(vii) B16 and Schulreich et al. (2017) used the epicycle approximation (Fuchs et al. 2006), 
which is less precise than integrating the gravitational potential.  \\
(viii) B16 determine the position of their predicted SNe by tracing back the 3D motion
of the current stars of UCL (and LCC) back to the time of each individual predicted SN,
and obtain the predicted SN position at that time from the highest space density of
currently existing stars in UCL (or LCC) at that time; this may constrain the position too much.

Regarding the total number of SNe expected in UCL and LCC, the different choices in B16 do not 
cancel each other, but the selected IMF slope, the neglected runaways and Black Holes,
and in particular the old association age (22.5 Myr) and the short lifetime of stars
with $\sim 8-12$ M$_{\odot}$ old star, all lead to more SNe and, hence, more energy input
into the Local Bubble.

For further discussion, see also Hyde \& Pecaut (2018) and Sorensen et al. (2017).
Our closest approach and SN 16 from B16 coincide in location, because both were within UCL at that time.
The fact that the B16 model seems to reproduce the measured $^{60}$Fe flux on Earth
regarding time distribution and amplitude may be conincidental (in particular regarding the
timing of SNe, see above), the absolute $^{60}$Fe flux is also uncertain -- B16:
``our model reproduces the measured {\it relative} abundances''. 

\section{A high-mass X-ray binary in LCC -- another SN that ejected $^{60}$Fe ?}

As mentioned above, depending on the total mass before and after the SN and the fraction of mass loss
on the total mass as well as the SN kick on the neutron star, the binary will either get unbound
(ejecting a runaway star) or it will remain bound.
In case of the latter, one could expect to observe a system comprised by a neutron star
and a normal star as High- or Low-Mass X-ray Binary (HMXB or LMXB, respectively),
if the separation is sufficiently small for accretion to occur.

We searched for both HMXB and LMXB in UpSco, UCL, and LCC in the catalogs
of Liu et al. (2000, 2007) and did find one HMXB in LCC (Table 6).

\begin{table}
\caption{Optical stars connected to the high-mass X-ray binary 1H11255-567 in LCC}
\begin{tabular}{l|l|l} \hline
Parameter & $\mu ^{2}$ Cru & $\mu ^{1}$ Cru \\ \hline
Spectral type  & B5Ve$^{a}$ & B2IV-Ve$^{b}$ \\
parallax [mas] & $8.95 \pm 0.23^{c}$ & (d) \\
V [mag] & 5.2$^{e}$ & 4.0$^{e}$ \\
mass [M$_{\odot}$] & $5 \pm 1^{f,g,h}$ & $7-8^{f,g}$ \\
RV [km/s] & $15.9 \pm 1.9^{i}$ & $14.4 \pm 0.6^{j}$ \\
Proj. rot. vel. $v \cdot \sin i$ [km/s] & $242 \pm 10^{k}$ & $34 \pm 3^{k}$ \\
Proper motion $\mu _{\alpha}$ [mas/yr] & -$28.16 \pm 0.22^{c}$ & (d) \\
Proper motion $\mu _{\delta}$ [mas/yr] & -$10.34 \pm 0.34^{c}$ & (d) \\ \hline
\end{tabular}

Remarks: (a) de Vaucouleurs (1957), (b) Hiltner et al. (1969), (c) Gaia DR2, 
(d) see text (Sect. 6) for Hipparcos and Gaia data,
(e) Ducati (2002),
(f) Tetzlaff et al. (2011a), 
(g) Hyde \& Pecaut (2018), 
(h) Adam (2017),
(i) mean of 19 radial velocity measurements (Jilinski et al. 2006),
(j) mean of 72 radial velocity measurements (Jilinski et al. 2006),
(k) Brown \& Verschueren (1997).

\end{table}

The {\it High Energy Astrophysical Observatory} (HEAO, Wood et al. 1984) 
X-ray source 1H11255-567 with $\mu ^{2}$ Cru was classified as HMXB, 
because the X-ray source is located at the position of the optical star $\mu ^{2}$ Cru,
which forms a wide common-proper-motion pair with $\mu ^{1}$ Cru (WDS, Mason et al. 2014).
In the last 20 Myr, 1H11255-567 was always within 8-12 pc of the center of LCC.
Given the spectral types of $\mu ^{2}$ and $\mu ^{1}$ Cru, it is dubious whether
they can produce X-ray emission on their own by colliding winds, coronal emission, or else.
The X-ray source was also detected by the
{\it Rossi X-ray Timing Explorer} (RXTE), see below. 
In the ROSAT All-Sky Survey (1990/91, 0.1-2.4 keV), 
neither $\mu ^{1}$ nor $\mu ^{2}$ Cru were detected with upper limits being
$\log$~L$_{X} \le 30.05$ (erg/s) for $\mu ^{1}$ Cru and $\log$~L$_{X} \le 29.89$ (erg/s) for $\mu ^{2}$ Cru 
(Bergh\"ofer \& Schmitt 1996).

Archival RXTE data (10.4 ks observed on 1998 Jan 13) were reduced by us: the source is to be identified with HMXB 1H11255-567;
the X-ray flux is $3 \cdot 10^{-12}$ erg/cm$^{2}$/s, the X-ray luminosity at the Gaia DR2 distance
of $\mu ^{2}$ Cru is then $5 \cdot 10^{30}$ erg/s;
the RXTE X-ray luminosity is a few orders of magnitude below the otherwise
X-ray faintest HMXB known, $\sim 10^{34}$ erg/s (Liu et al. 2000);
the RXTE X-ray spectrum can be fitted with XSPEC (reduced $\chi ^{2} = 0.7$ with 56 degrees of freedom)
for an absorbing photon model (phabs)
with negligible extinction
and a power law spectrum with high-energy cut-off;
a pure power law gives a similar good fit;
given that the spectrum shows only hard emission (above 2.4 keV), the X-ray emission is probably due to accretion.
As an X-ray emitting neutron star, the unresolved object would need to be
older, smaller, and/or cooler than, e.g., RXJ1856-3754.
If the separation between the X-ray source and the optical star $\mu ^{2}$ Cru is sufficiently
large, then accretion and, hence, X-ray flux can be weak.
The X-ray emission would also be consistent with an accreting Black Hole in the HMXB,
e.g. the HMXB MWC 656 with a Black Hole and a Be star has
$\log L_{X} \le 32$ (erg/s) (Casares et al. 2014), but Black Holes are rarer than neutron stars.

If there is a compact object, it may be less likely that it is a Black Hole,
because the mass function of
LCC hardly predicts one such massive progenitor -- e.g. according to B16,
the most massive star in LCC is expected to have had $\sim 19$ M$_{\odot}$,
close to the lowest suggested lower mass limit for Black Hole progenitors,
18~M$_{\odot}$ (Smartt 2015).
X-ray positional accuracy is not sufficient to determine the orbital separation
between this object and $\mu ^{2}$ Cru; the high rotational velocity of $\mu ^{2}$ Cru
might indicate (pre-SN) mass transfer.

We should also consider whether the X-ray source could be an unresolved low-mass companion
(spectral type G, K, or M),
which could emit X-rays due to accretion from a circumstellar disk
or due to its own magnetic activity.
At an age of 10-20 Myr like LCC, G- and K-type stars would be on the
zero-age main-sequence, while M-dwarfs would still be pre-main sequence.
For a bolometric luminosity of G-, K-, or M-dwarfs
(from solar mass stars to very low-mass late M-dwarfs) of 
$L_{\rm bol} = 0.003-1 \cdot L_{\odot}$ (e.g. Stelzer et al. 2016)
(and the X-ray luminosity as obtained above for our HMXB, $5 \cdot 10^{30}$ erg/s),
we would obtain $\log L_{\rm X}/L_{\rm bol} = -0.4$ to $-2.9$.
For young, low-mass M-dwarfs, this is above and, hence, incompatible with 
their saturation limit $\log L_{\rm X}/L_{\rm bol} \simeq -3$ 
(e.g. H\"unsch et al. 1999, Hambaryan et al. 2004, Stelzer et al. 2016).
Also, the X-ray saturation luminosity of late-type stars down to M6
is $\log L_{\rm X} = 28.7$ (erg/s) (Stelzer et al. 2016),
again below the X-ray luminosity of the HMXB.
Even lower-mass objects with spectral types L or T do not
emit X-rays at all (e.g. Stelzer et al. 2016).
Hence, a late M-dwarf cannot be responsible for the X-ray source.

G- and K-type zero-age main-sequence stars with 10-20 Myr typically have X-ray luminosities
in the range of $\log L_{\rm X} \simeq 29.5$ to 30 (erg/s), see e.g. Neuh\"auser et al. (1995),
a factor of 5 below the measured X-ray luminosity of what was classified a HMXB,
but a particularly active young G-/K-type star with $\sim 1$ M$_{\odot}$ may not be excluded.
For companions of such a star with semi-major axes $\ge 0.25$ au, the astrometric wobble 
for a circular orbit would be $\ge 1$ mas (i.e. detectable with Hipparcos and Gaia);
for orbital periods much larger than $\sim 20$ yr ($\sim 3$ au), the epoch difference between Hipparcos and Gaia,
the astrometric wobble is not yet detectable.
For semi-major axes below 0.25 au, the radial velocity signal would be $\sim 30$ km/s (for edge-on orbit inclination), 
i.e. well detectable -- but it is not observed, because the RV jitter seen in $\mu^{2}$ Cru is $\pm 1.9$ km/s only (Table 6);
the expected amplitude would reach the RV jitter seen in $\mu^{2}$ Cru ($\pm 1.9$ km/s) 
for an orbital inclination of $3.6^{\circ}$ only (almost face-on).
The system is also not known as eclipsing binary.
These limits leave only a small parameter range for an unresolved close ($\le 3$ au) low-mass companion,
but wider companions are not yet excluded.

Deep high-angular resolution IR imaging with Adaptive Optics has revealed a
possibly co-moving faint companion candidate $\sim 0.2^{\prime \prime}$ SE of $\mu^{2}$ Cru (22 au projected semi-major axis):
the system was observed with VLT-Naco --
first on 2005 Jan 27 (PI V. Ivanov, archival data) with the S27 camera (pixel scale $27.147 \pm 0.028^{\prime \prime}$)
with a total of 450s integration time (split in many small exposures) 
in the K$_{\rm s}$-band at an effective FWHM of 83 mas,
then again in 2nd epoch on 2011 May 9 (PI C. Adam) with the S13 camera (pixel scale $13.262 \pm 0.045^{\prime \prime}$)
with a total of 252s integration time (split in many small exposures),
again in the K$_{\rm s}$-band, now with an effective FWHM of 72 mas.
A faint point-like object is detected first $0.192 \pm 0.009^{\prime \prime}$ and then $0.204 \pm 0.006^{\prime \prime}$
off $\mu^{2}$ Cru with a position angle of $88.9 \pm 1.0^{\circ}$ and then $134.3 \pm 0.4^{\circ}$,
respectively, i.e. to the SE. 
While this change in position angle, compared to a non-moving background
object and considering the parallax and proper motion of $\mu^{2}$ Cru, can exclude a non-moving background and
can be consistent with a co-moving companion or a moving background object,
the negligible change in separation is inconclusively.
The magnitude difference between $\mu^{2}$ Cru ($5.337 \pm 0.021$ mag) 
and the faint companion candidate is $4.128 \pm 0.053$ mag in K$_{\rm s}$.
The data, an image, and more details about data reduction and astrometry can be found in Adam (2017).
For an age of 10-20 Myr and a Gaia DR2 distance of $111 \pm 3$ pc, 
using the models by Siess (2000), Bertelli et al. (2009), and Bressan et al. (2012),
one can obtain a mass for the faint object of $\sim 0.75 \pm 0.08$~M$_{\odot}$, if indeed a companion.
In principle, though, it is also possible that the faint object is an unrelated moving background object:
if it is another young member of the UCL association a few pc in the fore- or background of $\mu^{2}$ Cru,
then it could also have emitted the detected X-rays, but if it is an unrelated old object, then the X-ray
emission could come from an unresolved compact object.
To distinguish between a co-moving young companion or a moving old background object, one would need to obtain
more epochs (to detect curvature in orbital motion) or a spectrum (to detect evidence for youth).
If the faint object is an X-ray emitting companion to $\mu^{2}$ Cru, then this is not a HMXB,
and maybe a similar situation could be considered for other presumable high- and low-mass X-ray binaries.

Gaia DR2 data on parallax and proper motion of $\mu^{1}$ Cru are consistent with $\mu^{2}$ Cru
within $1.2 \sigma$, but the goodness of the astrometric fit of the Gaia DR2 data of $\mu^{1}$ Cru is not
sufficient, e.g. for inclusion in Simbad, and the error bars on the Gaia DR2 data of $\mu^{1}$ Cru are 
larger than those from Hipparcos -- possibly because the star is close to the Gaia brightness limit of 3.6 mag;
Gaia DR2 data of $\mu^{2}$ Cru are acceptable (Table 6), Gaia DR2 data of $\mu^{2}$ Cru are:
\begin{itemize}
\item Gaia $\mu^{2}$ Cru: $\pi = 9.63 \pm 0.36$ mas
\item Gaia $\mu^{2}$ Cru: $\mu _{\alpha} = -31.09 \pm 0.46$ mas/yr
\item Gaia $\mu^{2}$ Cru: $\mu _{\delta} = -14.51 \pm 0.50$ mas/yr
\end{itemize} 
The Hipparcos data (van Leeuwen 2007) on parallax and proper motion are: 
\begin{itemize}
\item Hipparcos $\mu^{1}$ Cru: $\pi = 7.87 \pm 0.17$ mas
\item Hipparcos $\mu^{1}$ Cru: $\mu _{\alpha} = -30.69 \pm 0.13$ mas/yr
\item Hipparcos $\mu^{1}$ Cru: $\mu _{\delta} = -13.08 \pm 0.11$ mas/yr

\smallskip

\item Hipparcos $\mu^{2}$ Cru:  $\pi = 8.01 \pm 0.29$ mas
\item Hipparcos $\mu^{2}$ Cru: $\mu _{\alpha} = -32.49 \pm 0.13$ mas/yr
\item Hipparcos $\mu^{2}$ Cru:$\mu _{\delta} = -10.92 \pm 0.19$ mas/yr
\end{itemize}
These parallaxes are consistent within less than $1 \sigma$.
The Hipparcos parallax of $\mu^{1}$ Cru is consistent with the Gaia DR2 parallax of $\mu^{1}$ Cru within $2 \sigma$. 
Both Hipparcos and Gaia data are fully consistent with orbital motion.
Also, according to the Washington Visual Double Star (WDS) catalog (Mason et al. 2014),
$\mu^{2}$ Cru is bound to $\mu^{1}$ Cru.

Let us assume for a moment that the X-ray source is a neutron star:
the sum of the masses of $\mu ^{1}$ Cru and $\mu ^{2}$ Cru plus $\sim 1.4$ M$_{\odot}$ for one 
neutron star (the X-ray source)
would give $\sim 14 \pm 0.5$ M$_{\odot}$; the total mass of the pre-SN multiple system must
have been less than twice this mass, so that it remained bound; hence, we obtain for the
SN progenitor a mass limit of $\le 15.6 \pm 0.5$ M$_{\odot}$ --
also, it must have been above the mass of $\mu ^{1}$ Cru, which did not explode, yet.
Therefore, the mass and spectral type of the neutron star progenitor could be 
intermediate between $\zeta$ Oph and the PSR\,B1706 progenitor
or lower in mass -- hence, lifetime intermediate or larger;
hence, the SN producing the X-ray source would have happened within the last $\sim 1.8$ Myr.
Tracing back the 3D motion of this X-ray binary, the pre-SN massive multiple star would have been
at a distance between 112 pc now and 89 pc (1.8 Myr ago),
so that this SN could also have contributed significantly to the $^{60}$Fe on Earth.
We display the motion of 1H11255-567 within LCC in Fig. 6.

The nature of the X-ray companion as neutron star, Black Hole, or low-mass companion
can be confirmed with high-resolution X-ray and/or infrared spectroscopic observations in the near future 
(e.g. a White Dwarf cannot have formed in LCC, yet, given its young age).

Since $^{60}$Fe is detected for the period $\sim 1.5-3.2$ Myr ago (Wallner et al. 2016),
and since the $^{60}$Fe travel time is $\sim 0.1-0.5$ Myr (Fry et al. 2015, B16), either this SN happened
$\sim 1.0-1.8$ Myr ago and the $^{60}$Fe arrived on Earth $\sim 1.5-2.3$ Myr ago;
or the SN took place $1-5 \cdot  10^{5}$ yr ago, so that -- depending on the $^{60}$Fe travel time (Fry et al. 2015, B16) --
$^{60}$Fe has not yet arrived on Earth (and more than $\sim 10^{5}$ yr ago, so that a SN remnant does not exist any more).
If all the $^{60}$Fe, which arrived $\sim 1.5-2.3$ Myr ago, can be explained with just the one SN producing PSR\,B1706-16,
then the SN producing the HMXB exploded $1-5 \cdot 10^{5}$ yr ago (or otherwise did not contribute to the $^{60}$Fe).
If the HMXB does contain a Black Hole with 2-10 M$_{\odot}$,
the mass range of Black Holes in X-ray binaries (Ritter \& Kolb 2015), then the system also would have remained bound --
for a somewhat higher progenitor mass and shorter lifetime (maybe without SN and without $^{60}$Fe).

The considerations about this HMXB remain preliminary until the nature of the X-ray source (compact or low-mass) is confirmed.

\begin{figure*}
\centering
\includegraphics[width=16.6cm]{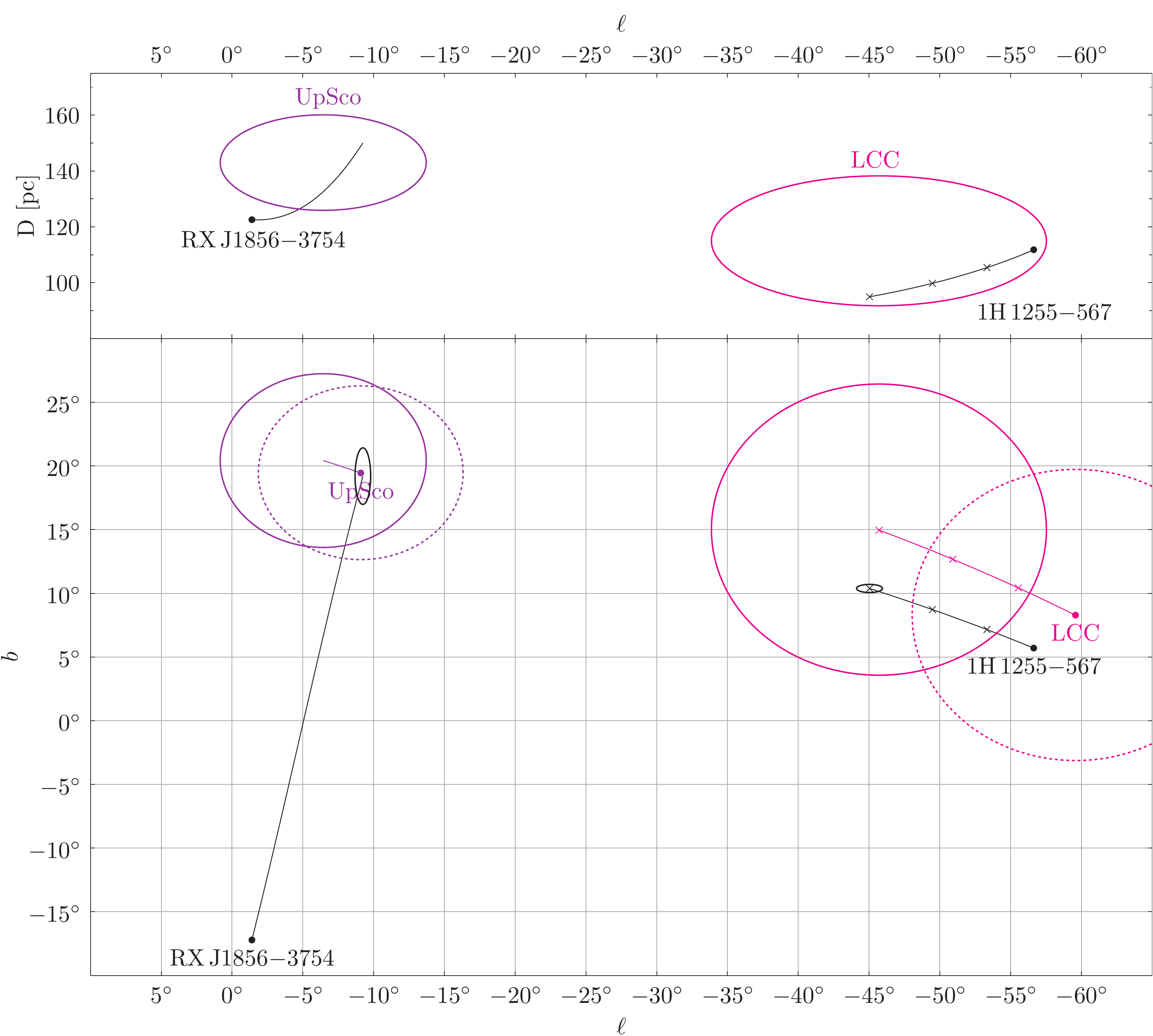}
\caption{The high-mass X-ray binary 1H11255-567 in LCC and the isolated neutron star RXJ\,1856-3754
coming from UpSco may both have formed in SNe in Scorpius-Centaurus-Lupus. The runaway-pulsar-pair
$\zeta$ Oph and PSR\,B1706-16 are not shown here for clarity (see Fig. 2).
Right part: the motion of the high-mass X-ray binary 1H11255-567 in LCC,
where its possible neutron star formed in a SN within the last 1.8 Myr (Sect. 6).
In the lower right, Galactic longitude l versus latitude b,
the black dot is its location now (112 pc), the black line its path on sky for the last 1.5 Myr (then at 89 pc)
with tick mark crosses after every 0.5 Myr.
The ellipse around its position 1.5 Myr ago indicates the uncertainty.
The small red dot and red dotted circle show the center and extent of LCC now, 
the full red circle its current extend 1.5 Myr ago shifted by its bulk proper motion; 
the red line shows its path on sky with tick mark crosses after every 0.5 Myr.
Left part: the motion of the isolated neutron star RXJ\,1856-3754 within the last 0.5 Myr coming out of UpSco
(Tetzlaff et al. 2010).
In the lower left, Galactic longitude l versus latitude b,
the black dot is its location now (123 pc), the black line its path on sky for the last 0.5 Myr.
The small pink dot and pink dotted circle show the center and extent of UpSco now,
the full pink circle its current extend 0.5 Myr ago shifted by its bulk proper motion; 
the pink line shows its path on sky in the last 0.5 Myr.
The upper panel shows the third dimension: Galactic longitude l versus distance D (in pc).
}
\end{figure*}

\section{Discussion}

After preparing and testing our software to trace back the motion of stars
through the Galactic potential (Sect. 2), we revised the work by Ho00 and Ho01 
by showing that $\zeta$ Oph and pulsar PSR\,B1929+10 cannot have had 
a common origin (their only case of a pulsar-runaway-pair), see Sect. 3,
but that instead, $\zeta$ Oph and the radio pulsar PSR\,B1706-16 may well
have been at the same time at the same location (inside UCL),
where a SN in a binary should have released them (Sect. 4).
With new Gaia runaway stars and additional runaway-pulsar-systems, 
the frequency of SN ejection of runaway stars can be studied in the near future.

We also concluded that the $^{60}$Fe found on Earth can indeed be linked to a certain SN 
by tracing back runaway and neutron stars: 
$1.78 \pm 0.21$ Myr ago, a SN in UCL from a primary star with at least 16 and at most 18-19 M$_{\odot}$ 
at $107 \pm 4$ pc ejected its companion $\zeta$ Oph as runaway, formed the pulsar PSR\,B1706-16,
and delivered $^{60}$Fe to Earth.

The observed $^{60}$Fe flux in the Earth ocean crust $F_{\rm obs}$ is related to the distance $d$ 
and the time $t$ of the SN (i.e. the travel time of $^{60}$Fe from the SN to Earth) by
\begin{equation}
F_{\rm obs} = \frac{1}{4} \cdot \frac{M_{ej}}{4 \cdot \pi \cdot d^{2} \cdot 60 \cdot m_{\rm u}} \cdot U \cdot e^{-t/\tau}
\end{equation}
with the $^{60}$Fe yield $M_{\rm ej}$ ejected by the SN 
(atomic mass number $A=60$ for $^{60}$Fe, $m_{\rm u}$ for atomic mass unit), 
the travel time $t$ of the $^{60}$Fe, 
the uptake factor $U$ as fraction of the isotope that arrives at Earth in the form of dust being 
incorporated into the Earth ocean crust 
(anywhere from $> 0$ to $\le 1$), 
and the mean lifetime $\tau = 2.6$ Myr of $^{60}$Fe (e.g. Fry et al. 2015).
We could constrain distance $d$ and time $t$ of the SN,
but $U$ and the $^{60}$Fe yield of SNe remain unknown.
If other isotopes could be detected, the uptake $U$ can be constrained.

The $^{60}$Fe yield of SNe depends on the progenitor mass (e.g. Rauscher et al. 2002 for theoretical models),
but would also need to be constrained by observations.
If we can identify a neutron star ejected in a SN that brought $^{60}$Fe to Earth,
and if we can trace it back to the location of the SN (e.g. with a runaway star),
we would know not only the distance towards the SN and its exact time,
but also the lifetime of the progenitor (association age minus flight time) and, hence, its mass
(see, e.g., Neuh\"auser et al. 2012).
Then, one can constrain uptake and/or SN yield.

A revised expectation from the UCL mass function for the {\it consensus age} of $16 \pm 2$ Myr (Hyde \& Pecaut 2018) 
expects that a SN progenitor with $\sim 14-16$ M$_{\odot}$ exploded $\sim 1-3$ Myr ago to 
deliver $^{60}$Fe to Earth (Hyde \& Pecaut 2018), consistent with our scenario. 

In another theoretical work (Fry et al. 2015), the $^{60}$Fe on Earth is linked to a core-collapse SN 
from a $\sim 15$ M$_{\odot}$ progenitor at $\sim 94^{+19}_{-12}$ pc with $^{60}$Fe travel time 
of $\sim 0.44$ Myr (or $\sim 0.50$ Myr for $\sim 107$ pc); 
then, our SN scenario $1.78 \pm 0.21$ Myr ago agrees well with the observed peak $^{60}$Fe at $\sim 2.2$ Myr. 
That model by Fry et al. (2015) is for certain $^{60}$Fe yields (Rauscher et al. 2002) and U=0.5 as ocean crust uptake factor; 
once the latter is fixed (e.g. by another isotope), SN yields can be constrained.
Alternatively, since different SN yield models are similar at $\sim 16$ M$_{\odot}$ 
(Rauscher et al. 2002, Fry et al. 2015, Wallner et al. 2016, B16), 
the agreement of theoretical expectations (Fry et al. 2015) combined with those yields (Rauscher et al. 2002) 
for a SN at $\sim 94$ pc with our empirical scenario ($\sim 107$ pc) 
for $\sim 16$ M$_{\odot}$ would constrain the uptake factor to U$\simeq 0.65$ (using $U \propto 1/d^{2}$ from Equ. 5).

While in Fry et al. (2015) and other publications, an isotropic distribution of $^{60}$Fe on sky and homogeneous arrival at Earth
was assumed, Fry et al. (2016) considered a non-isotropic distribution and deflection of grains by the
Earth magnetic field, resulting in a smaller amount that arrived -- or in a smaller SN distance needed
for the same observed flux:
e.g., a core-collapse SN from a 15 M$_{\odot}$ star with U=0.5 would need to be at a distance of $64^{+14}_{-8}$ pc
to provide all the detected $^{60}$Fe (Fry et al. 2016).
While the directional information for $^{60}$Fe on Earth is mostly lost due to atmospheric and oceanic circularization,
such information should be available for $^{60}$Fe found on the Moon (Firmani et al. 2016):
Fry et al. (2016) found {\it that it is not yet possible to differentiate between a Sco-Cen or Tuc-Hor source
because the sample were drawn near the lunar equator and the large uncertainties in the measurements}.

While Fry et al. (2015, 2016) tried to explain the whole $^{60}$Fe peak (on Earth) at $\sim 2.2$ Myr with 
just one SN, Wallner et al. (2016) showed that the dispersed $^{60}$Fe arrival from 1.5 to 3.2 Myr can be explained
only with several SNe (their figure S6) -- plus at least one more for the $\sim 8$ Myr old extra signal in
Wallner et al. (2016). 
Since Fry et al. (2015, 2016) considered only one SN for the origin on the whole $^{60}$Fe in their
model, we consider any constraint on SN yields or the $^{60}$Fe uptake (with their model results, see above) as preliminary.
If there were several SNe instead if just one, each of the individual SNe would need
to bring less $^{60}$Fe than the one considered in Fry et al. (2015, 2016) -- and for similar $^{60}$Fe yields
and progenitor masses, they can then be at larger distance than expected by Fry et al. (2015, 2016). 
Our evidence for one such SN (Sect. 4) is then consistent with having contributed to
the $^{60}$Fe detected on Earth.

In addition, the only high-mass X-ray binary in Scorpius-Centaurus-Lupus
(1H11255-567 with $\mu^{1}$ and $\mu^{2}$ Cru) may include a neutron star formed in another SN,
which should have exploded up to 1.8 Myr ago 
located at 89-112 pc, i.e. also yielding $^{60}$Fe detectable on Earth (Sect. 6, Fig. 6). 

We noted that the X-ray bright radio-quiet neutron star RXJ\,1856.5-3754 may also have originated 
in Scorpius-Centaurus-Lupus, namely in UpSco $\sim 0.5$ Myr ago at a distance of $\sim 150$ pc
(Tetzlaff et al. 2010), the only OB association, through which its path on sky did pass
in the last 3.8 Myr (its characteristic spin-down upper age limit, van Kerkwijk \& Kaplan 2008) 
-- but this was probably too recent and/or too distant for detection of $^{60}$Fe from its SN on Earth.
See also Fig. 6.

In the near future, we will also consider all other stellar associations suggested for recent nearby SNe
(see Pecaut \& Mamajek 2018, Sorenson et al. 2017) -- and after the final Gaia data release, also many new Gaia runaway stars.

The connection between $^{60}$Fe on Earth and nearby recent SNe offers advances regarding nuclear physics (nucleosynthesis), 
interstellar medium ($^{60}$Fe transport time and losses), cosmic-ray and helio-physics (injection into the solar system), 
geophysics ($^{60}$Fe uptake in ocean crust, tentatively constrained here to $\sim 0.65$), 
astrobiology (perturbation of Earth biosphere by enhanced cosmic-ray activity) (Fields et al. 2019) 
as well as astrophysics (star and planet formation trigger and feedback due to SNe).
Given the distance towards the SN (or even two SNe) found here to be just outside of 100 pc
(instead of within 100 pc as considered earlier, e.g. Melott 2016), one should now re-consider 
possible effects on the Earth biosphere and evolution.

\section{Summary of results}

We presented the following work and results:
\begin{itemize}
\item We wrote software to trace back stars and associations in the Galactic potential for a few Myr,
tested successfully with an original case from Ho00 and Ho01, runaway star $\zeta$ Oph and pulsar PSR\,B1929+10.
\item We could then show with their most recent data (e.g. revised RV of $\zeta$ Oph from Zehe et al. 2018)
that $\zeta$ Oph and PSR\,B1929+10 do not have a common origin.
\item We then compared to past flight paths of all ten well-known OB-type runaway stars from
or near Sco-Cen-Lup (Ho00, Ho01) 
and all 400 Galactic neutron stars with known distance and transverse velocity (Manchester et al. 2005)
to the Sco-Cen-Lup association.
\item We found only one runaway-pulsar-pair that satisfies all our criteria for a close meeting
in the past, namely runaway star $\zeta$ Oph and radio pulsar PSR\,B1706-16:
within less than 0.5 pc, they were at the same time ($1.78 \pm 0.21$ Myr ago) at the same distance ($107 \pm 4$ pc).
It is therefore most likely, that at this location and time, a SN in a binary took place releasing both
the runaway star and the pulsar.
Our conclusions hold for a RV of the pulsar of $260 \pm 43$ km/s -- and we would like to stress
that with newly found runaway stars and neutron stars as well as more accurate kinematic data,
modifications may arise in the future.
\item From the difference between association age and flight time, we can then estimate the life-time
and, hence, mass of the SN progenitor: 16-18 M$_{\odot}$.
\item Given the time and distance of this SN, it may have contributed to $^{60}$Fe detected on Earth.
In such a case, it may constraint $^{60}$Fe uptake in the Earth crust and/or SN nucleosynthesis yields.
\item Given the current flight directions and velocities, the neutron star PSR\,B1706-16
obtained a kick of $253 \pm 54$ km/s during the SN.
\item The two runaway stars $\kappa ^{1}$ Aps (HIP 76013) and j Cen (HIP 57669) also come from Sco-Cen-Lup,
so that they may have been companions to SN progenitors, which released $^{60}$Fe in their SNe
(2.9-0.5 Myr ago) --  but without any neutron stars found at their respective positions in the past.
\item Within Sco-Cen-Lup, there is only one low-mass or high-mass X-ray binary known,
namely 1H11255-567 with $\mu^{1}$ and $\mu^{2}$ Cru; this system emits hard X-rays.
If the unresoled X-ray emitting companion
is a neutron star, it would have formed in a SN, which should have exploded up to 1.8 Myr ago and
located at 89-112 pc, so that it may also have contributed to the $^{60}$Fe detected on Earth
(however, the neutron star nature of this object is uncertain).
\end{itemize}

\bigskip

\noindent {\bf Acknowledgements.}
RN had the idea for this project in 2008 when reading page 85 in the book 
{\it Das geschenkte Universum} by Arnold Benz and would like to thank Dagmar L. Neuh\"auser 
for long-term support and many good questions and suggestions. 
We are also grateful for discussions with Dieter Breitschwerdt; 
we also thank Markus Mugrauer and Oliver Lux for help and advice. 
We acknowledge Christian Adam for the material on the faint companion candidate near $\mu^{2}$ Cru.
We acknowledge financial support from the Deutsche Forschungsgemeinschaft in grant NE 515/59-1. 
This work has made use of data from the European Space Agency missions Hipparcos and Gaia. 
We also used the ATNF pulsar catalog at www.atnf.csiro.au (Manchester et al. 2005). 
We also thank an anonymous referee for many valuable comments.

\end{document}